\emergencystretch=\maxdimen
\documentclass[12pt,a4paper]{article} % A4 paper and 11pt font size

\usepackage[utf8]{inputenc} % Polskie znaki. Potrzebne tylko w niektórych kompilatorach?
\usepackage[english]{babel}
\usepackage{mathtools,amsmath,amsfonts,amsthm,amssymb} % Matematyczne symbole, czcionki itp.
\numberwithin{equation}{section} % Numerowanie równań wewnatrz sekcji.
\usepackage{authblk} % Specyfikacja afiliacji.
\usepackage{indentfirst} % Wcinanie pierwszych paragrafów.
\usepackage{sectsty} % Komenda ponizej.
\allsectionsfont{\centering \normalfont \scshape} % Tytuły sekcji na srodku, ładna czcionka.
\usepackage[margin=2cm]{geometry} % Zmniejszenie marginesów do 3cm.
\usepackage{color}
\usepackage{tikz-cd}

\usepackage{sidecap}
\sidecaptionvpos{figure}{c}

\newcommand{\I}{\mathrm{i}}

\title{Dynamics of a lattice $2$-group gauge theory model}
\author{A. Bochniak, L. Hadasz, P. Korcyl and B. Ruba}
\affil{\textit{Institute of Theoretical Physics, Jagiellonian University}}
\affil{\textit{prof. Łojasiewicza 11, 30-348 Kraków, Poland}}
\date{\today}

\begin{document}
\frenchspacing

\maketitle

\begin{abstract}
We study a simple lattice model with local symmetry, whose construction is based on a crossed module of finite groups. Its dynamical degrees of freedom are associated both to links and faces of a four-dimensional lattice. In special limits the discussed model reduces to certain known topological quantum field theories. In this work we focus on its dynamics, which we study both analytically and using Monte Carlo simulations. We prove a factorization theorem which reduces computation of correlation functions of local observables to known, simpler models. This, combined with standard Krammers-Wannier type dualities, allows us to propose a detailed phase diagram, which form is then confirmed in numerical simulations. We describe also topological charges present in the model, its symmetries and symmetry breaking patterns. The corresponding order parameters are the Polyakov loop and its generalization, which we call a Polyakov surface. The latter is particularly interesting, as it is beyond the scope of the factorization theorem. As shown by the numerical results, expectation value of Polyakov surface may serve to detects all phase transitions and is sensitive to a value of the topological charge. 
\end{abstract}

%It turns out that its expe

%Certain limits of this model can be identified with certain types of topological quantum field theories. In the selected model, independent degrees of freedom are associated to both links and faces of a four-dimensional lattice and are subject to a certain constraint. We present the details of this construction and obtain general results concerning the expected dynamics. We identify four regions of phase space with different phases. We provide numerical evidence for our statements from  Monte Carlo simulations. In order to describe the phase diagram of the model we propose two pairs of local and non-local order parameters. In particular, we propose a generalization of Wilson and Polyakov lines to Wilson and Polyakov surfaces. We provide theoretical expectations for their behaviour in different phases which we confirm through numerical results.

\section{Introduction}

Higher gauge theories are physical models which generalize conventional gauge theory by associating degrees of freedom to geometric objects of dimension higher than one. Perhaps the best known example is the $p$-form electrodynamics \cite{pform}, whose discretized version, a natural generalization of the Wilson formulation of ordinary gauge theory \cite{savit, orland}, can be  expressed in terms of degrees of freedom associated to $p$-cells, e.g.\ plaquettes for $p=2$. These degrees of freedom are subject to redundancy described by group valued functions on the set of $(p-1)$-cells. In the case of $p=1$ this reduces to degrees of freedom on links with gauge transformations given by arbitrary functions defined on lattice sites.  

Already in \cite{pform} it was argued that gauge theories with $p \geq 2$ are necessarily abelian, essentially because there exist no well-behaved orderings on surfaces. There exists a~way to bypass this argument, inspired by higher category theory \cite{baez, pfeiffer, baez_huerta}. For $p$ not exceeding $2$, it~is typically formulated in terms of $2$-groups \cite{baez1} or, equivalently, crossed modules \cite{brown}. Surface observables in $2$-group gauge theories are still valued in an abelian group, but in general they are computed in terms of genuinely non-abelian local degrees of freedom associated to links and plaquettes.

There exists also a concept of (global) higher form symmetries \cite{global}, whose relation with higher gauge theories is similar to the relation between ordinary symmetries and gauge theories. Examples of models admitting higher symmetries have been known for a long time, and  among gauge theories they are in fact the rule rather than an exception. Nevertheless, systematic study of higher symmetries seems to have begun relatively recently.

Higher gauge theories have been proposed \cite{gukov, KT14} as effective field theories describing vacua of conventional gauge theories. They also provide interesting examples \cite{kapth, yetter,porter, martins_porter, gpp08, williamson} of Topological Quantum Field Theories (TQFTs) \cite{atiyah, witten}, and hence are expected to describe certain gapped topological phases of many body quantum systems. In \cite{kapth} the existence of Symmetry Protected Topological (SPT) phases protected by higher symmetries was proposed. Another motivation to study higher gauge theories is provided by its relation with bosonization in arbitrary dimension \cite{bos1,bos2,bos3}. Furthermore, certain models in string theory may be described as higher gauge theories \cite{saemann}.

Yetter's model \cite{yetter} is a TQFT based on a crossed module of finite groups. Its hamiltonian realizations resembling the Kitaev's toric code were constructed in \cite{bullivant17, delcamp, bullivant20}. In \cite{bhr} a common generalization of the Yetter's model, $2$-form $\mathbb Z_n$ electrodynamics and lattice Yang-Mills theory has been proposed. It is a genuinely dynamical model, formulated in the hamiltonian formalism, which reduces to a TQFT only in certain limits. In this work we consider an analogous model formulated in terms of state sums (or discrete functional integrals). We focus on one relatively simple crossed module, but some of our results are true in general. In order to make the paper more accessible, we have decided to define everything explicitly using notations standard in lattice gauge theory. We refer to \cite{bhr} for an exposition of the slightly more involved formalism of crossed modules and proofs of various algebraic facts used in this text.

%to include local physical degrees of freedom. This model is a common generalization of Yetter's model, $2$-form $\mathbb Z_n$ electrodynamics and lattice Yang-Mills theory. 

Full definition of the model under consideration is given in subsection \ref{DOF}. Its extended observables are discussed in subsection \ref{order}. We identify topological charges and higher symmetries: $1$-form symmetry $\mathbb Z_2^{(1)}$ and $2$-form symmetry $\mathbb Z_2^{(2)}$. We discuss the theoretical possibility of symmetry breaking and provide suitable order parameters. In subsection \ref{reduction} we show that computation of a large class of observables, including all local observables, may be reduced to calculation of averages in simpler models: $1$-form $\mathbb Z_2$ gauge theory and $2$-form $\mathbb Z_2$ gauge theory. This includes the statement that plaquette observables (constructed from link degrees of freedom in the usual way) are uncorrelated with cube observables (constructed from plaquette degrees of freedom), which is not obvious from the form of the action. This factorization theorem is not valid for the surface observable which is the order parameter of the $\mathbb Z_2^{(2)}$ symmetry. In subsection \ref{phase_diag} we use the factorization theorem and Kramers-Wannier type dualities to formulate a proposal for the phase diagram. We describe critical points, symmetry breaking patterns and renormalization group fixed points governing the infrared physics.

Section \ref{sec:MC} is devoted to Monte Carlo study of the proposed model in dimension $D=4$. Simulation algorithm is described in subsection \ref{sec:MC1}. Since the general model is fairly standard, we discuss in detail only those aspects that are specific to the case at hand. In~subsection \ref{sec:MC2} we present numerical results for expectation values of local observables. These results confirm the phase structure obtained from duality arguments. The most interesting, in our view, results of simulations are presented in subsection \ref{sec:MC3}. They concern behaviour of order parameters for higher symmetries $\mathbb Z_2^{(1)}$ and $\mathbb Z_2^{(2)}$. It is found that order parameters for the latter not only exhibit sharp dependence on both coupling constants, but also are sensitive to the topological charge. 

Finally, in the appendix \ref{app:surface}, we discuss construction of non-spherical surface observables using the general language of crossed modules.

\section{Description of the model}

\subsection{Degrees of freedom, action and gauge freedom} \label{DOF}

Coordinates of a lattice site form a tuple $x=(x_1,...,x_D)$ with integer $x_{\mu}$. Unit vector in the direction $\mu  \in \{ 1,...,D\}$ will be denoted by $\widehat \mu$. We choose periodic boundary conditions, i.e. $x_{\mu}$ is identified with $x_{\mu} + L_{\mu}$, where $L_{\mu}$ is the spatial extent of the system in the $\mu$-th direction. 

\begin{SCfigure}
  \centering
  \includegraphics[width=0.4\textwidth]{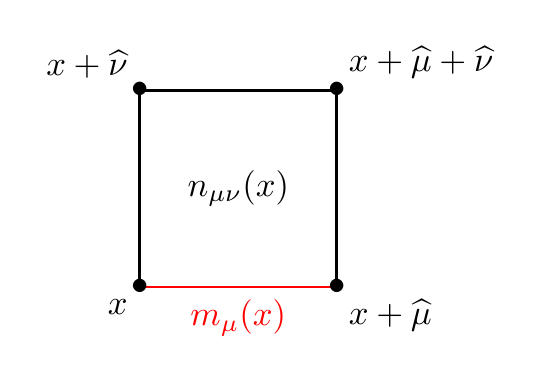}  
  \caption{Independent degrees of freedom are associated to links $m_{\mu}(x)$ and faces $n_{\mu \nu}(x)$. The latter should not be confused with plaquette observables $f_{\mu \nu}(x)$ constructed from links.}
  \label{fig:deg_plaq}
\end{SCfigure}

Addition and multiplication in $\mathbb Z_4 = \{ 0,1,2,3\}$ is always performed modulo four. We consider a~model with degrees of freedom of two types, both valued in $\mathbb Z_4$ (see Fig. \ref{fig:deg_plaq}):
\begin{itemize}
    \item $m_{\mu}(x)$, associated with the link between $x$ and $x + \widehat \mu$,
    \item $n_{\mu \nu}(x)$, associated with the square with corners $x, \, x + \widehat \mu, \, x + \widehat \mu + \widehat \nu, x + \widehat \nu$ (called face).
\end{itemize}

They are subject to a constraint (for every $x$ and $\mu < \nu$) called \emph{fake flatness}:
\begin{equation}
2 n_{\mu \nu}(x) = m_{\mu}(x) + m_{\nu}(x+\widehat \mu) - m_{\mu}(x + \widehat \nu) - m_{\nu}(x).
\label{eq:fake_flatness}
\end{equation}
The right hand side is a plaquette built of $m$ variables as in the ordinary lattice gauge theory. It is convenient to denote it by $f_{\mu \nu}(x)$. We note that $n_{\mu \nu}(x)$ is determined by the link variables only modulo two and that $f_{\mu \nu}(x)$ has to be even (but $m_{\mu}(x)$ not necessarily so).

Out of elementary degrees of freedom one may construct observables associated to cubes:
\begin{subequations} \label{eq:cube_def}
\begin{align}
g_{\mu \nu \rho}(x) &= -n_{\mu \nu}(x) +n_{\mu \rho}(x) - n_{\nu \rho}(x)  \tag{\ref{eq:cube_def}} \\
  & + (-1)^{m_{\rho}(x)} n_{\mu \nu}(x+ \widehat \rho) - (-1)^{m_{\nu}(x)} n_{\mu \rho}(x+ \widehat \nu)  + (-1)^{m_{\mu}(x)} n_{\nu \rho}(x+ \widehat \mu) \nonumber.
\end{align}
\end{subequations}
The six terms in this formula correspond to six sides of a cube. It can be shown that fake flatness enforces all $g_{\mu \nu \rho}(x)$ to be even. 

Observable $f_{\mu \nu}(x)$ is the Wilson line along the boundary of an elementary rectangle. In the present model it is possible to construct also higher dimensional analogues of Wilson lines, which could be called Wilson surfaces. Observable $g_{\mu \nu \rho}(x)$ is the Wilson surface along an elementary cube.

We choose the following action:
\begin{equation} \label{eq:action}
S = - J_1 \sum_{x} \sum_{\mu < \nu} (-1)^{\frac{f_{\mu \nu}(x)}{2}} - J_2 \sum_{x} \sum_{\mu < \nu < \rho} (-1)^{\frac{g_{\mu \nu \rho}(x)}{2}} = J_1 S_1(m) + J_2 S_2(m,n),
\end{equation}
with $J_1, J_2 \geq 0$. The first term is the Wilson action for $m$ variables. It is minimized if all plaquettes $f_{\mu \nu}(x)$ are equal to zero. Every plaquette equal to $2$ costs $2J_1$ units of action. The second term is a higher dimensional analogue of the Wilson term for the $n$ variables. Again, it is minimized if all cubes $g_{\mu \nu \rho}(x)$ are equal to zero. Every excited cube costs $2 J_2$ units of action.

Degrees of freedom superficially seem to interact with each other, since they are related by the fake flatness condition and since $g_{\mu \nu \rho}(x)$ (which enters the action directly) depends on both degrees of freedom. However, as it will be demonstrated later, this interaction does not  affect local dynamics, i.e. plaquettes are uncorrelated with cubes and furthermore correlation functions of plaquettes and cubes depend only on $J_1$ and only on $J_2$, respectively. On the other hand, the impact of the interaction can be seen in averages of nonlocal order parameters. Numerical evidence supporting this statement is presented in section \ref{sec:MC}.

%These statements have supporting numerical evidence described in Section \ref{sec:MC}}.

The fake flatness constraint \eqref{eq:fake_flatness} and the action \eqref{eq:action} are invariant under gauge transformations of two types. Gauge transformations associated to sites are parametrized by elements $\xi(x) \in \mathbb Z_4$. They act according to the formulas:
\begin{subequations}
\begin{gather}
m_{\mu}(x) \mapsto m_{\mu}(x) + \xi(x + \widehat \mu) - \xi(x), \\
n_{\mu \nu}(x) \mapsto (-1)^{\xi(x)} n_{\mu \nu}(x).
\end{gather}
\end{subequations}
Gauge transformations associated to links are parametrized by $\psi_{\mu}(x) \in \{ 0, 2 \} \subsetneq \mathbb Z_4$ and act as
\begin{subequations} \label{eq:edge_gauge_trans}
\begin{gather}
m_{\mu}(x) \mapsto m_{\mu}(x), \\
n_{\mu \nu}(x) \mapsto n_{\mu \nu}(x) + \psi_{\mu}(x) + \psi_{\nu}(x + \widehat \mu) - \psi_{\mu}(x + \widehat \nu) - \psi_{\nu}(x). \label{eq:edge_gauge_transf_b}
\end{gather}
\label{eq:edge_gauge_transf}
\end{subequations}
Only gauge invariant quantities will be regarded as observables. In this work we consider $f_{\mu \nu}(x)$, $g_{\mu \nu \rho}(x)$ and order parameters described in subsection \ref{order}. 

\subsection{Nonlocal order parameters and symmetries} \label{order}

Polyakov loop, a particular Wilson line winding around one of the directions of the lattice, is defined by the formula
\begin{equation}
\label{eq:polyakov line}
p_{\mu}(x) = \exp \left(\frac{ \I \pi}{2}{\sum \limits_{j=0}^{L_{\mu}-1} m_{\mu}(x + j \widehat \mu)}\right).
\end{equation}
Its possible values are $\pm 1$ and $\pm \I$, in contrast to plaquette observables which take only two possible values. For configurations with $f_{\mu \nu}(x)=0$ for all $x$, value of $p_{\mu}(x)$ is independent of~$x$. This is not true for general configurations. On the other hand, quantity $Q_{\mu}$ defined by
\begin{equation}
\label{eq:topological charge}
    Q_{\mu} = p_{\mu}(x)^2
\end{equation} 
is independent of $x$, which follows from the fact that all $f_{\mu \nu}(x)$ are even. We will call it the topological charge. Each $Q_{\mu}$ may take two possible values, $1$ or $-1$, so the whole set of field configurations decomposes into $2^D$ disjoint sectors. We note that local constraint-preserving transformations in the set of all field configurations cannot change the topological charge, since that requires changing $p_{\mu}(x)$ for all $x$.

For every $\mu$ there exists a symmetry of the action which leaves all plaquettes, cubes and $\{ p_{\nu}(x) \}_{\nu \neq \mu}$ unchanged, but flips the sign of $p_{\mu}(x)$ (and hence preserves $Q_{\mu}$). It is given by
\begin{subequations}
\begin{gather}
 m_{\nu}(x)  \mapsto m_{\nu}(x) + 2 \delta_{\mu, \nu} \delta_{x_{\mu},0}, \\
 n_{\nu \rho}(x)  \mapsto n_{\nu \rho}(x).
\end{gather}
\label{eq:1form_sym}
\end{subequations}
We will call it the electric $1$-form symmetry. As a consequence of this symmetry the expectation value of $p_{\mu}(x)$ vanishes. 

We are unaware of a symmetry which changes $p_{\mu}(x)$ by a factor of $\I$ (and hence flips the sign of $Q_{\mu}$). Nevertheless, it will turn out to be useful to consider the $Q_{\mu}$-reversing transformation
\begin{subequations}
\begin{gather}
 m_{\nu}(x)  \mapsto m_{\nu}(x) + \delta_{\mu, \nu} \delta_{x_{\mu},0}, \\
 n_{\nu \rho}(x)  \mapsto n_{\nu \rho}(x).
\end{gather}
\label{eq:Q_trans}
\end{subequations}
It preserves fake flatness and all $f_{\mu \nu}(x)$, so it is a symmetry of $S_1$. However, it changes values of cube observables, so it is not a symmetry of the full action.

We would like to address the question whether the symmetry \eqref{eq:1form_sym} can be spontaneously broken. We insist on gauge invariance and locality of the action, so it is not possible to include a symmetry breaking term in the action. On the other hand, in a putative phase with unbroken symmetry the infinite volume limit of the volume average of $p_{\mu}(x)$ in a fixed typical field configuration is expected to vanish. This happens for small $J_1$, because then plaquette observables fluctuate strongly, so the signs of $p_{\mu}(x)$ and $p_{\mu}(y)$ are essentially independent if the transverse distance $|x-y|_{\perp} = \sqrt{\sum \limits_{\nu \neq \mu} (x_{\nu} - y_{\nu})^2}$ is large. More precisely, $p_{\mu}(x) p_{\mu}(y)^{-1}$ may be understood as a Wilson loop bounding area $L_{\mu}|x-y|_{\perp}$, so its expectation value is expected to decay exponentially with $|x-y|_{\perp}$.

To quantify the above discussion, we consider
\begin{equation}
 P_{\mu} = \left| V_{\perp}^{-1} \sum_{x} p_{\mu}(x) \right|,
 \label{eq:pmu_avg}
\end{equation}
with the sum taken over $x$ in a plane transverse to the $\mu$-th direction. Here $V_{\perp} = \prod \limits_{\nu \neq \mu} L_{\nu}$ is the transverse volume. Squaring this definition we find
\begin{equation}
P_{\mu}^2 = V_{\perp}^{-2} \sum_{x,y} p_{\mu}(x) p_{\mu}(y)^{-1}.
\end{equation}
There are $V_{\perp}^2$ terms in this sum, each of which has modulus one. After taking expectation value, only $O(V_{\perp})$ terms, with $|x-y|_{\perp}$ comparable to the correlation length survive. Therefore the average of $P_{\mu}^2$ decreases as $V_{\perp}^{-1}$, so $P_{\mu}$ decreases as $V_{\perp}^{- \frac{1}{2}}$:
\begin{equation}
    P_{\mu} \sim V_{\perp}^{- \frac{1}{2}}, \qquad L_{\mu} \text{ fixed, } V_{\perp} \to \infty.
\end{equation}
By spontaneous breaking of the symmetry~\eqref{eq:1form_sym} we shall understand violation of this scaling law, so that $P_{\mu}$ remains nonzero in the limit of infinite transverse volume:
\begin{equation}
    P_{\mu} \sim \mathrm{const.} \neq 0, \qquad L_{\mu} \text{ fixed, } V_{\perp} \to \infty.
\end{equation}
Note that it may still be true that $P_{\mu} \to 0$ as $L_{\mu} \to \infty$.

There exists a surface observable analogous to the Polyakov loop. It may be thought of as a~Wilson surface winding around two lattice directions. Its construction is slightly more involved. We choose a plane through a fixed site $x$ parallel to directions $\mu < \nu$. Morally speaking, we would like to add $n_{\mu \nu}(y)$ with $y$ running through all sites in the chosen plane. However, this does not give a gauge invariant quantity. To fix this, we have to choose for every $y$ a path from $y$ to $x$ (which we take to be entirely contained in the chosen plane) and weigh $n_{\mu \nu}(y)$ by a parallel transport factor $\prod (-1)^{m_{\rho}(z)}$, where the product is taken over all links forming the chosen path. Then the sum, denoted $\Sigma_{\mu \nu}(x)$, is gauge-invariant and even. This is discussed in the broader context of crossed modules in the appendix \ref{app:surface}. We define
\begin{equation}
\label{eq:polyakov plane}
p_{\mu \nu}(x) = \exp \left( \frac{\I \pi}{2} \Sigma_{\mu \nu}(x) \right),
\end{equation}
which will be called the Polyakov plane.

We remark that our notation is fully justified only if either $Q_{\mu}=Q_{\nu}=1$ or all $f_{\mu \nu}(x)$ vanish, because otherwise $p_{\mu \nu}(x)$ depends on the choice of paths needed to construct it. This hints at the possibility that expectation values of $p_{\mu \nu}(x)$ may depend both on the two coupling constants and on the topological charge. This will be corroborated by results in section \ref{sec:MC}.

There exists a symmetry which flips the sign of $p_{\mu \nu}(x)$:
\begin{subequations}
\begin{align}
    m_{\rho}(x) & \mapsto m_{\rho}(x), \\
    n_{\rho \sigma }(x) & \mapsto n_{\rho \sigma}(x) + 2 \delta_{\mu, \rho} \delta_{\nu, \sigma} \delta_{x_{\mu},0} \delta_{x_{\nu},0} , \qquad \qquad \rho < \sigma. 
\end{align}
\label{eq:2form_sym}
\end{subequations}
We will call it electric $2$-form symmetry. It implies that the expectation value of $p_{\mu \nu}(x)$ vanishes.

By analogy with the Polyakov loop, we consider the quantity
\begin{equation}
P_{\mu \nu} = \left| V_{\perp}^{-1} \sum_x p_{\mu \nu}(x) \right|,
\label{eq:pmunu_avg}
\end{equation}
where $V_{\perp} = \prod \limits_{\rho \neq \mu, \nu} L_{\rho}$ and the sum is taken over a plane transverse to $\widehat \mu$ and $\widehat \nu$. We will say that the symmetry \eqref{eq:2form_sym} is broken if $P_{\mu \nu}$ has nonzero limit as $V_{\perp} \to \infty$.

\subsection{Reduction of dynamics to simpler models} \label{reduction}

In this subsection we will show how to express certain averages with respect to the action \eqref{eq:action} in terms of averages in simpler models. We will make use of constraint-preserving moves in the space of field configurations. Firstly, the link moves:
\begin{subequations}
\begin{align}
m_{\mu}(x) & \mapsto m_{\mu}(x) + 2 \psi_{\mu}(x), \label{eq:m_psi_move} \\
n_{\mu \nu}(x) & \mapsto n_{\mu \nu}(x) + (-1)^{m_{\mu}(x)} \psi_{\mu}(x) +(-1)^{m_{\mu}(x + \widehat \nu) + m_{\nu}(x)} \psi_{\nu}(x + \widehat \mu)  \\
 &- (-1)^{m_{\mu}(x + \widehat \nu) + m_{\nu}(x)} \psi_{\mu}(x + \widehat \nu) - (-1)^{m_{\nu}(x)} \psi_{\nu}(x), \nonumber
\end{align}
\end{subequations}
with arbitrary $\psi_{\mu}(x) \in \mathbb Z_4$. They reduce to gauge transformations \eqref{eq:edge_gauge_trans} if $\psi_{\mu}(x)$ is even. In general they change the value of plaquette observables $f_{\mu \nu}(x)$, but not of the cube observables $g_{\mu \nu \rho}(x)$. Secondly, the face moves:
\begin{subequations}
\begin{gather}
\label{eq:n_chi_move}
m_{\mu}(x) \mapsto m_{\mu}(x), \\
n_{\mu \nu}(x) \mapsto n_{\mu \nu}(x) + \chi_{\mu \nu}(x),
\end{gather}
\end{subequations}
with $\chi_{\mu \nu}(x) \in \{ 0 , 2 \} \subsetneq \mathbb Z_4$. They preserve $f_{\mu \nu}(x)$, but change values of $g_{\mu \nu \rho}(x)$.

Moves described above generate a group. Every move may be represented as a sequence of~local moves with only one nonzero $\psi_{\mu}(x)$ or $\chi_{\mu \nu}(x)$. Since $m_{\mu}(x)$ are always either unchanged or shifted by an even amount, topological charges $Q_{\mu}$ are invariant. 

We claim that any two configurations with equal topological charges can be related by a~sequence of local moves and a gauge transformation. Indeed, first consider two configurations with equal $m_{\mu}(x)$. Then, by fake flatness, all $n_{\mu \nu}(x)$ differ by even numbers, so~the two configurations are related by a face transformation. This reduces the proof of the claim to showing that $m_{\mu}(x)$ can be made equal by a sequence of link moves and a gauge transformation. The only gauge invariant functions of $m_{\mu}(x)$ are Wilson lines, which can be taken along contractible loops or non-contractible loops. The former are expressible in terms of $f_{\mu \nu}(x)$ and have to be even. The latter are also even on the account of the assumption about topological charges, since every loop can be built of contractible loops and Polyakov loops. This proves that up to pure gauge terms, the difference of $m_{\mu}(x)$ is even. Thus they are related by a transformation of the form \eqref{eq:m_psi_move}.

The average of an observable $O$ is $\langle O \rangle = \frac{Z_O(J_1,J_2)}{Z(J_1,J_2)}$, where $Z(J_1,J_2) = Z_{\mathbf 1}(J_1, J_2)$ and
\begin{equation}
Z_O(J_1,J_2) = \sum_{m,n} f(Q) e^{- J_1 S_1(m) - J_2 S_2(m,n)} O(m,n),
\end{equation}
in which the sum over $m,n$ is restricted by the constraint. Function $f(Q)$ is a weight given to the sector with topological charge $Q$. The simplest choice is $f(Q)=1$, while restriction to $Q=Q'$ with fixed $Q'$ corresponds to $f(Q) = \delta_{Q,Q'}$. We consider an observable of the form $O= O_1 O_2$ such that:
\begin{itemize}
    \item $O_1$ can by expressed solely in terms of plaquette observables $f_{\mu \nu}(x)$ (thus it can be $P_{\mu}$),
    \item $O_2$ is invariant with respect to gauge transformations and link moves, e.g.\ it is an arbitrary function of cube observables $g_{\mu \nu \rho}(x)$.
\end{itemize}

We define the quantity
\begin{equation}
    W_{O_2}(J_2;m) = \sum_n e^{-J_2 S_2(m,n)} O_2(m,n).
    \label{eq:W_def}
\end{equation}
It is invariant with respect to gauge transformations and link moves of $m$ variables, so it depends on $m$ only through $Q_{\mu}$. Therefore we write $W_{O_2}(J_2;m)= W_{O_2,Q}(J_2)$, which gives
\begin{equation}
Z_{O_1 O_2}(J_1,J_2) = \sum_m f(Q) e^{-J_1 S_1(m)} O_1(m)W_{O_2,Q}(J_2).
\end{equation}
We divide the summation over $m$ into topological sectors. The sum over $m$ with fixed $Q$ will be denoted by index $m | Q$:
\begin{equation}
Z_{O_1 O_2}(J_1,J_2) = \sum_Q f(Q) W_{O_2,Q}(J_2) \sum_{m | Q } e^{- J_1 S_1(m)} O_1(m).
\end{equation}
Sum $\sum\limits_{m | Q } e^{- J_1 S_1(m)} O_1(m)$ does not depend on $Q$ by symmetry \eqref{eq:Q_trans} of $S_1$. Finally:
\begin{equation}
Z_{O_1 O_2}(J_1,J_2) = \left( \sum_{m | 1} e^{-J_1 S_1(m)} O_1(m) \right) \left( \sum_Q f(Q) W_{O_2,Q}(J_2) \right).
\end{equation}
In the remaining sum over $m$ we have configurations of $m$ variables such that every Wilson loops is even. Such configuration is gauge equivalent to one with all $m_{\mu}(x)$ even. Furthermore, every gauge orbit has $4^{N_1-1}$ representatives (in which $N_1$ is the number of links), out of which $2^{N_1-1}$ is such that all $m_{\mu}(x)$ are even. Therefore we may restrict the sum over $m$ to configurations with even $m_{\mu}(x)$ at the small cost of including a factor $2^{N_1-1}$. Then the sum over $m$ gives the Wegner model \cite{wegner}, so
\begin{equation}
Z_{O_1 O_2}(J_1,J_2) = 2^{N_1-1} Z^{\mathrm{Wegner}}_{O_1}(J_1) \sum_Q f(Q) W_{O_2,Q}(J_2).
\label{eq:factorization}
\end{equation}
This gives
\begin{equation}
\langle O_1 O_2 \rangle = \frac{Z_{O_1 O_2}(J_1,J_2)}{Z(J_1,J_2)} = \frac{Z^{\mathrm{Wegner}}_{O_1}(J_1)}{Z^{\mathrm{Wegner}}(J_1)} \cdot \frac{\sum\limits_Q f(Q) W_{O_2,Q}(J_2)}{\sum\limits_Q f(Q) W_{\mathbf{1},Q}(J_2)},
\label{eq:factorization2}
\end{equation}
from which we can draw the following conclusions:
\begin{itemize}
    \item factorization $\langle O_1 O_2 \rangle = \langle O_1 \rangle \langle O_2 \rangle $,
    \item $\langle O_1 \rangle$ does not depend on $J_2$ and weights $f$, and is equal to the average in Wegner's model,
    \item $\langle O_2 \rangle $ does not depend on $J_1$.
\end{itemize}
This factorization theorem is the main result of this section. We would like to remark that its derivation remains valid also for models based on general crossed modules of finite groups, as long as the action is a sum of a term depending only on plaquette observables and a term depending only on cube observables. This observation follows from the fact that the presented proof relies only on general properties of gauge transformations and constraint-preserving moves. These were discussed in detail in \cite{bhr}.

Next we turn to the question on how $\langle O_2 \rangle$ depends on the topological charge sector. We will argue that for thermodynamic quantities dependence becomes negligible in the infinite volume limit. This will be confirmed already for quite small lattices by results of simulations presented in section \ref{sec:MC}.

Consider, for concreteness, the case $Q_1=-1$, $Q_{\mu} =1$ for $\mu \neq 1$. Such choice of topological charge may be realized by the gauge field
\begin{equation}
    m_{\mu}(x) = \delta_{\mu, 1} \delta_{x_{\mu},0}.
\end{equation}
It is supported on a plane, so switching it on (without modifying $n_{\mu \nu}(x)$ variables) may change the value of at most $\binom{D-1}{2} \prod \limits_{\mu \neq 1} L_{\mu}$ cubes. Hence we have
\begin{equation}
    | S_2(m,n) - S_2(0,n)| \leq 2 \binom{D-1}{2} \prod_{\mu \neq 1} L_{\mu}.
\end{equation}
It follows that $\frac{W_{\mathbf{1},Q}(J_2)}{W_{\mathbf{1},\mathrm{trivial}}(J_2)} = \frac{\sum\limits_n e^{- J_2 S_2(0,n)} e^{- J_2 (S_2(m,n)-S_2(0,n))}}{\sum\limits_n e^{- J_2 S_2(0,n)}}$ obeys an estimate
\begin{equation}
 e^{-4 J_2 \binom{D-1}{2} \prod_{\mu \neq 1} L_{\mu}}  \leq  \frac{W_{\mathbf{1},Q}(J_2)}{W_{\mathbf{1},\mathrm{trivial}}(J_2)} \leq e^{4 J_2 \binom{D-1}{2} \prod_{\mu \neq 1} L_{\mu}}.
\end{equation}
Taking logarithms gives an estimate on the difference of free energies per unit volume:
\begin{equation}
\left|  \frac{1}{J_2} \log (W_{\mathbf{1},Q}(J_2))  - \frac{1}{J_2} \log (W_{\mathbf{1},\mathrm{trivial}}(J_2))  \right| \leq 4 \binom{D-1}{2} \prod_{\mu \neq 1} L_{\mu}.
\end{equation}
We recall that the free energy is an extensive quantity. On the other hand, the right hand side divided by the volume decays as $L_1^{-1}$ as $L_1 \to \infty$. We conclude that in the infinite volume limit, the free energies per unit volume become equal in all topological sectors.

\subsection{Phase diagram proposal for $D=4$} \label{phase_diag}

In this subsection we restrict attention to dimension $D=4$, although some parts of the discussion are valid also for other dimensions.

In the case $D=4$, Wegner's model has a~single phase transition \cite{wegner}, which is of first order. Its exact position
\begin{equation}
J_1^{\mathrm{crit}} = \frac{1}{2} \mathrm{arsinh}(1) \approx 0.441
\label{eq:j1 critical value}
\end{equation}
may be calculated using Kramers-Wannier type self-duality\footnote{Strictly speaking, the duality relates the partition function of the Wegner's model to the partition function summed over flat background $2$-form $\mathbb Z_2$ gauge fields. However, these gauge fields have negligible effect on thermodynamic quantities, which can be shown analogously as in the last paragraph of subsection \ref{reduction}.} \cite{kw}. There exist two renormalization group fixed points at $J_1 =0$ and $J_1 = \infty$. Two phases may be interpreted as their basins of attraction. 

At the point $J_1=0$, degrees of freedom become completely random and hence the theory is trivial. Effect of a small, but nonzero $J_1$ may be calculated using the strong coupling expansion \cite{wegner_flow}. One finds that Wilson loops obey the area law, and hence the electric $1$-form symmetry is unbroken. 

At $J_1 = \infty$ the system is constrained to configurations which minimize the action. Thus all plaquette observables vanish and Polyakov loops become independent of position. Up to gauge transformations, minima of the action are labeled by values of Polyakov loops. In Wegner's model each $P_{\mu}$ takes $2$ possible values, so there exist $16$ minima. They all have the same value of the action, because they are connected by the electric $1$-form symmetry. However, in order for the system to get from one minimum to another using local moves only, it has to overcome an infinite action barrier. Even for finite $J_1$ (but large, so that a typical configuration is close to a~minimum) the height of the barrier is of order $J_1 V_{\perp}$, so one may expect the electric $1$-form symmetry to be broken. 

The link variable sector of our model is slightly different in that the Polyakov loop takes four, rather than two possible values. However, it becomes essentially equivalent to the Wegner's model after restricting to a single topological charge sector.

Next we turn to the local dynamics of plaquette degrees of freedom. There exists a Kramers-Wannier duality between $W_{\mathbf{1}, \mathrm{trivial}}(J_2)$ and the Ising model partition function\footnote{Again, this is exact only if the Ising model partition function is summed over background $\mathbb Z_2$ gauge fields.} with
\begin{equation}
    \sinh(2J_2)  \sinh(2 J_{\mathrm{Ising}}) = 1.
    \label{eq. j2 critical}
\end{equation}
In the Ising model one expects a single continuous phase transition\footnote{We remark that in \cite{weak} a weakly first order phase transition was suggested.} whose position reported in \cite{lundow} is $J_{\mathrm{Ising}}^{\mathrm{crit}}=0.149647(5)$. This corresponds to a continuous phase transition in our model at \begin{equation}
J_{2}^{\mathrm{crit}}=0.953294(1).
\label{eq:j2 critical value}
\end{equation}
The critical point of the Ising model is expected to be described by a massless scalar field theory. It admits one relevant perturbation, given by the mass term. Therefore the fixed point at $J_2 = J_2^{\mathrm{crit}}$ is repulsive. The only other fixed points at $J_2=0$ and $J_2 = \infty$ describe physics in phases $J_2 < J_2^{\mathrm{crit}}$ and $J_2 > J_2^{\mathrm{crit}}$, respectively. 

Quite analogously to the Wegner's model, the electric $2$-form symmetry is unbroken in the small $J_2$ phase. The situation is much more interesting for large $J_2$. To gain some orientation about this case, we consider the limit $J_2 = \infty$, in which configurations are constrained to minimize $S_2$. As shown in the appendix \ref{app:surface}, Polyakov surfaces $p_{\mu \nu}(x)$ become independent of~$x$ if in addition either $J_1 = \infty$ (i.e.\ for configurations minimizing also $S_1$) or if topological charges are trivial. Therefore we expect that the $2$-form symmetry is broken if $J_2 > J_2^{\mathrm{crit}}$ and $J_1 > J_1^{\mathrm{crit}}$. In~the phase $J_2 > J_2^{\mathrm{crit}}$, $J_1 < J_1^{\mathrm{crit}}$ we can make this conclusion only for the sector with trivial topological charge. On the other hand, numerical results in section \ref{sec:MC} show that in the sector with $Q_{\mu}=-1$ the symmetry is restored. We find this result quite striking.

%and either $J_1 > J_1^{\mathrm{crit}}$ or $J_1^{\mathrm{crit}}$

%Then spontaneous breaking of the electric $2$-form symmetry is in principle possible. The symmetry could, however, be~restored by sufficiently strong fluctuations of the link variables (after all, in the definition of $P_{\mu \nu}(x)$ we were forced to use parallel transports constructed of link variables). These exist in the small $J_1$ phase, but not for large $J_1$.

%In the limit $J_1 = \infty$ only configurations with $f_{\mu \nu}(x)=0$ contribute. Then all $n_{\mu \nu}(x)$ are even, and~therefore $n_{\mu \nu}(x) = - n_{\mu \nu}(x)$. This means that parallel transport factors in the definition of $P_{\mu \nu}(x)$ become unimportant. Based on this observation we propose that for large $J_2$ and $J_1$, the electric $2$-form symmetry is broken, regardless of the value of the topological charge. The situation is less clear for $J_1=0$.

%We note that there exists a~qualitative difference between sectors in which it is true or not true that $Q_{\mu}=Q_{\nu}=1$. In the former case $P_{\mu \nu}(x)$ is actually independent of the choice of parallel transport paths, which indicates that dependence on $J_1$ may be milder. This is in accord with numerical results in section \ref{sec:MC}, which show that for large $J_2$, small $J_1$ and $Q_{\mu}=Q_{\nu}=1$ the symmetry is broken, but it becomes restored for $Q_{\mu}=-1$. We find this result quite striking.

The following picture emerges. Our model has four phases, each corresponding to one attractive renormalization group fixed point. In each of the fixed points local dynamics becomes trivial, but some nonlocal observables remain important:
\begin{itemize}
    \item $(J_1, J_2)=(0,0)$: $\mathbb Z_2$ topological charges $Q_{\mu}$,
    \item $(J_1, J_2) = (\infty,0)$: $\mathbb Z_4$ Polyakov loops $P_{\mu}$,
    \item $(J_1, J_2) = (\infty, \infty)$: $\mathbb Z_4$ Polyakov loops $P_{\mu}$ and $\mathbb Z_2$ Polyakov surfaces $P_{\mu \nu}$, completely independent of each other,
    \item $(J_1, J_2) = (0, \infty)$: $\mathbb Z_4$ Polyakov loops $P_{\mu}$ and $\mathbb Z_2$ Polyakov surfaces $P_{\mu \nu}$, with an interplay between topological charges and Polyakov surfaces.
\end{itemize}
We remark that the four renormalization group fixed points described here may be identified with four integrable hamiltonians described in \cite{bhr}.

\section{Monte Carlo study} \label{sec:MC}

\subsection{Simulation method} \label{sec:MC1}

%we consider the model on a four-dimensional torus (with periodic boundary conditions in each direction).

In the numerical setup, we keep the extent of three directions equal $L_0 = L_1 = L_2 = L$, whereas $L_3$ will be varied separately. We denote the entire volume by $V = L^3 \times L_3$. We will also use the notation
\begin{equation}
(x_0, x_1, x_2, x_3) = (x,y,z,t).
\end{equation}

For any observable we define its statistical expectation value, denoted by $\langle \cdot \rangle$, as the arithmetic mean over samples from a single Markov chain, and in some cases a weighted average of expectation values from multiple Markov chains. In most cases we perform a~single simulation where we gather around $10^5$ measurements{\color{blue},} from which we estimate the average and its standard deviation{\color{blue},} taking into account autocorrelations. We do the latter by explicitly calculating the autocorrelation function and integrating it up to the first non-positive element to quantify the autocorrelation time $\tau_{\textrm{int}}$. In~the following figures{\color{blue},} all data points are shown together with their statistical uncertanties, which however may be smaller than the symbol size and hence invisible. In some cases we have performed up to four parallel simulations in order to increase the statistics and to check for ergodicity.

All simulations are performed using an intertwined application of Metropolis \cite{metropolis,metropolis1,metropolis2} and over-relaxation steps \cite{overrelaxation,overrelaxation1,overrelaxation2,overrelaxation3,overrelaxation4}. These are two independent update steps coming in pairs: one for updating the link variables and another to update face variables. We now describe both in more details.

Metropolis steps are based on local changes separately for both kind of degrees of freedom. We use \eqref{eq:m_psi_move} to update the link variables and \eqref{eq:n_chi_move} for the face variables. We remind that by construction such moves preserve the fake-flatness constraint. As a consequence, the move \eqref{eq:m_psi_move} changes both link and face variables. If the constraint was satisfied by the initial configuration, it will be satisfied during the successive application of any of the above changes. Any two configurations can be linked by a finite-length chain of such local movements{\color{blue},} which ensures that the simulations are ergodic. Each new configuration $\nu$ is obtained from a previous configuration $\mu$ by a local change of a randomly chosen degree of freedom and is subject to an accept/reject step with a probability given by
\begin{equation}
\label{eq:accept/reject step}
    p_A(\mu \rightarrow \nu) = \min \big\{ 1, e^{ S(\mu) - S(\nu)}\big\}.
\end{equation}

%Over-relaxation steps are made of non-local moves which change the orientation of Polyakov lines $P_{\mu}$ Eq.~\eqref{eq:1form_sym} and Polyakov planes $P_{\mu \nu}$ Eq.~\eqref{eq:2form_sym}. 

Over-relaxation steps are made of non-local moves \eqref{eq:1form_sym} and \eqref{eq:2form_sym}, which flip the signs of Polyakov lines $P_{\mu}$ and Polyakov planes $P_{\mu \nu}$, respectively. Since such transformations do not change the value of the action of a given configuration, they would be always accepted. Hence they are not subject to the accept/reject step. It is known that the incorporation of such moves between Metropolis moves reduces autocorrelation times significantly \cite{overrelaxation,overrelaxation1,overrelaxation2,overrelaxation3,overrelaxation4}.

%drastically speeds up the thermalization of the Markov chain and hence reduces autocorrelation times \cite{overrelaxation}.

The local moves \eqref{eq:m_psi_move} and \eqref{eq:n_chi_move} cannot change the value of the topological charge. Hence, the simulation is limited to the topological sector given by the value of the topological charge of the initial configuration. In the following we discuss two independent chains of simulations, one performed in the trivial topological sector ($Q_{\mu} = 1$ for all $\mu$) and the second performed in the sector with $Q_0 = -1$, see \eqref{eq:topological charge}. We construct the latter by starting from an initial configuration where all the link and face variables are set to 0. Subsequently we set $m_{0}(0,y,z,t) = 1$ for all $y,z,t$, thus enforcing $P_0 = \I$ and hence $Q_0=-1$. 

The above algorithm with the accept/reject as in \eqref{eq:accept/reject step} satisfies the detailed balance condition, which together with the ergodicity of the local moves, guarantees the correctness of the entire algorithm in a given topological sector.

In order to identify the thermalization region of the Markov chain we usually perform an additional simulation with the same parameters{\color{blue},} which we start from a so-called hot initial configuration. The latter is constructed by randomizing as much as possible all the degrees of freedom. To be more precise{\color{blue},} we set $m_{\mu}(x,y,z,t)$ to 0 or 2 with equal probabilities, and subsequently adjust $n_{\mu \nu}(x,y,z,t)$ variables to satisfy the fake-flatness constraint. We do this by evaluating all plaquette variables $f_{\mu \nu}(x,y,z,t)$ and then setting $n_{\mu \nu} = \frac{1}{2}f_{\mu \nu}(x,y,z,t) + q$, where $q$ is a random variable taking values 0 and 2 with equal probability. In both simulations, started from a cold and hot configuration, we monitor two local variables \eqref{eq:average plaquette} and \eqref{eq:average cubes}. Recording of relevant observables is started only when the two monitored quantities attain compatible values.

% and start measuring relevant observables

\subsection{Numerical results for local observables} \label{sec:MC2}

%Results to be demonstrated:
%\begin{itemize}
    %\item Plaquettes first order transition (hysteresis plot), independent of $J_2$ and %topological charge
    %\item Cubes second order transition (continuous action, peak of fluctuations) %independent of $J_1$ and topological charge
%\end{itemize}

\begin{figure}
  \centering
  \includegraphics[width=0.345\textwidth, angle=270]{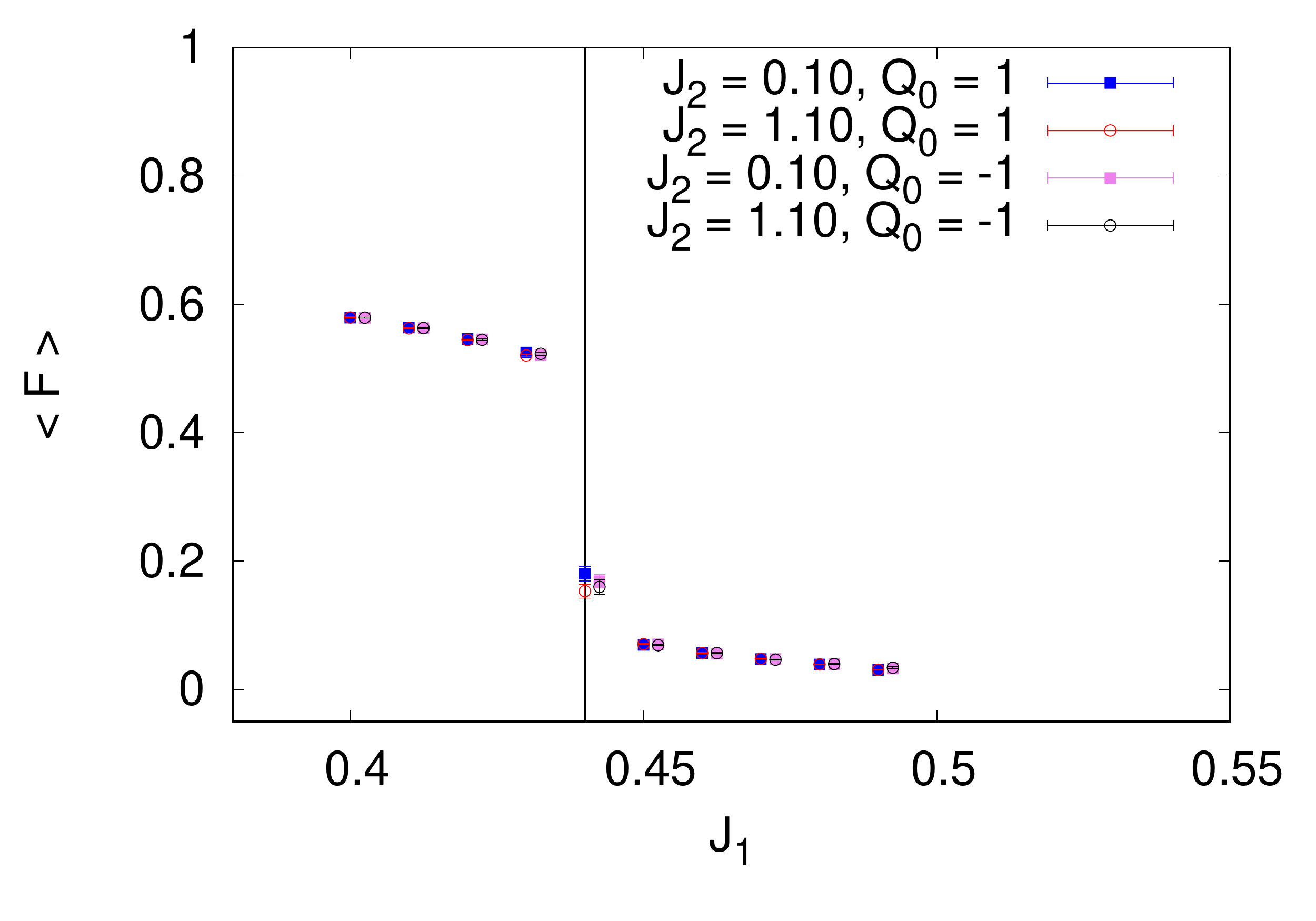}  
  \includegraphics[width=0.345\textwidth, angle=270]{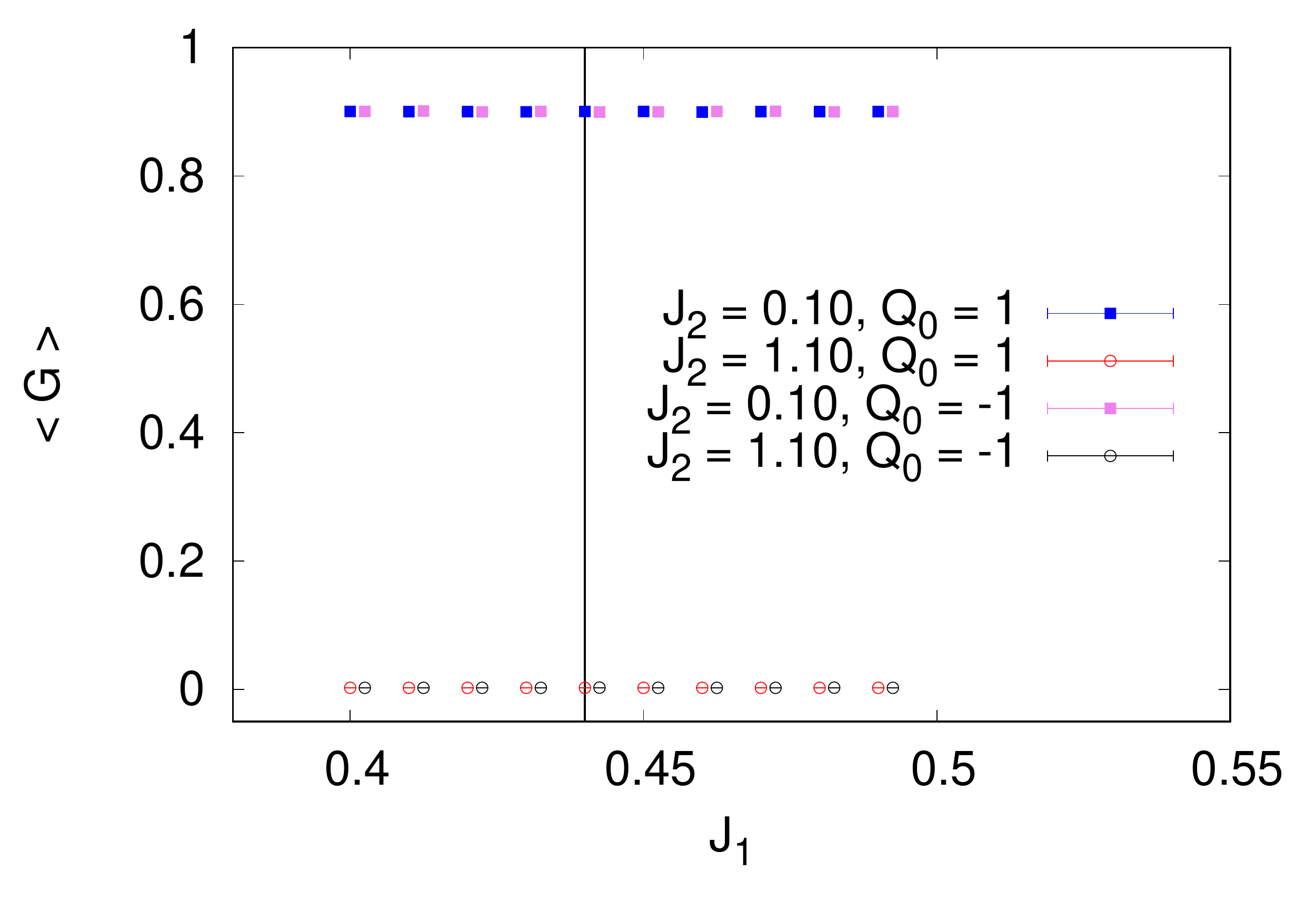}  
  \includegraphics[width=0.345\textwidth, angle=270]{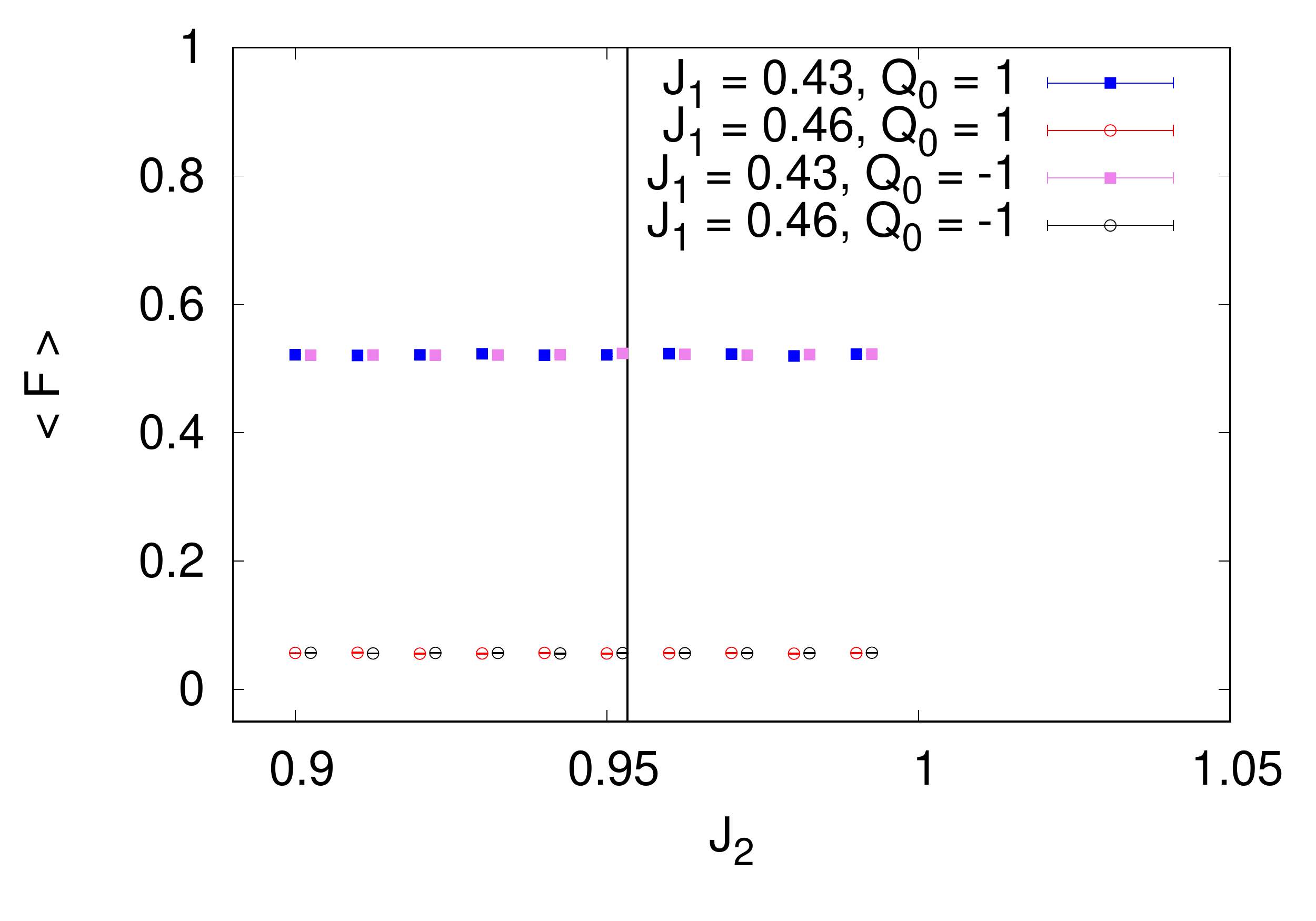}  
  \includegraphics[width=0.345\textwidth, angle=270]{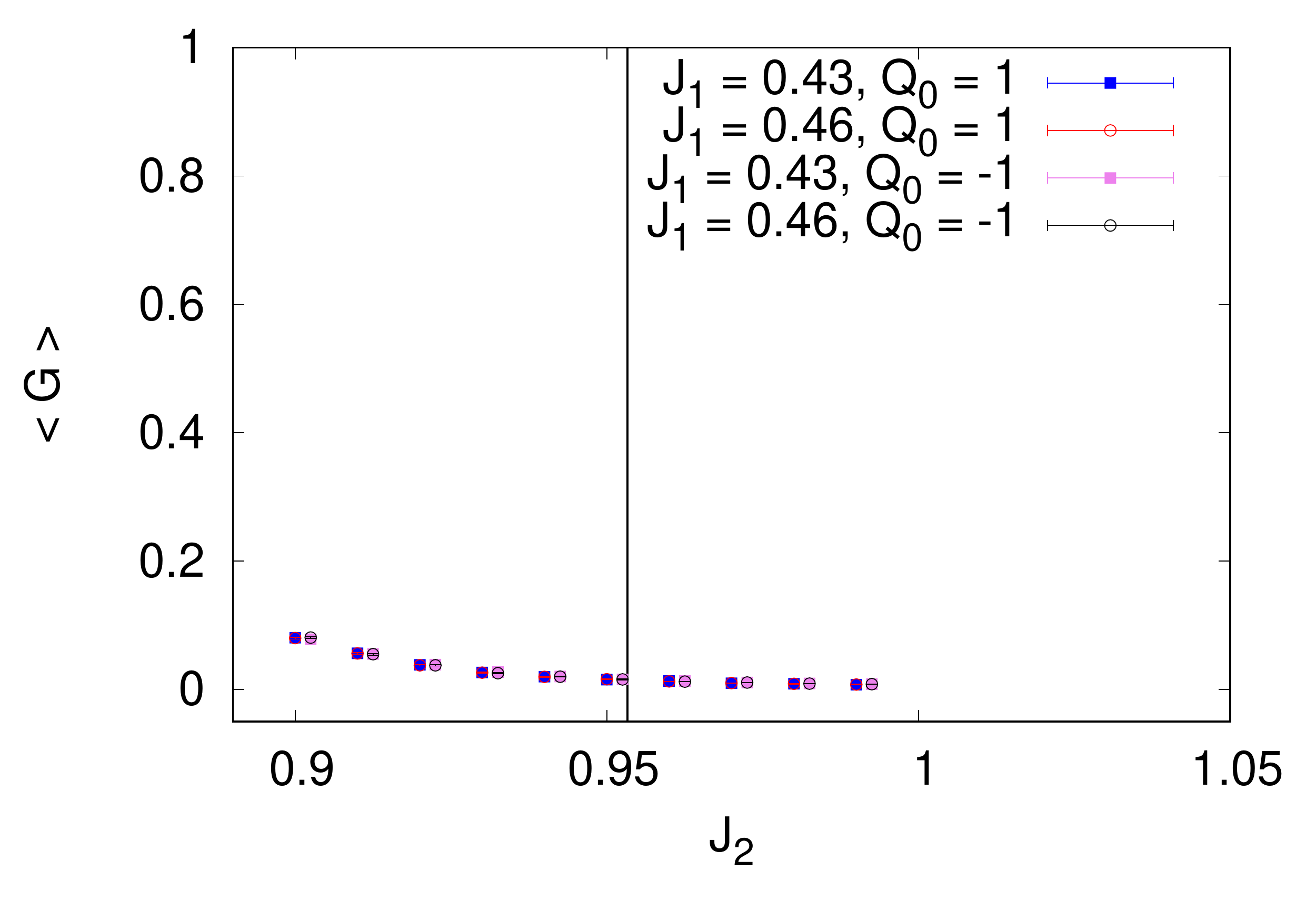}  
  \caption{Dependence of the plaquette and cube observables on $J_1$ (upper row) and $J_2$ (lower row) coupling constants. As predicted by the factorization theorem, $\langle G \rangle$ does not depend on $J_1$ (upper right panel), whereas $\langle F \rangle$ does not depend on $J_2$ (lower left panel). $\langle F \rangle$ show a significant jump around the expected first order phase transition marked by the solid vertical black line on the upper, right panel. As far as $\langle G \rangle$ is concerned, lower right panel shows a rather smooth dependence and no significant signs of the expected second order phase transition marked again by the solid vertical line. Fig.~\ref{fig:fluctuations and histeresis} demonstrates that indeed a second order phase transition happens around the expected $J^{\textrm{crit}}_2$. In all the cases, results from both topological sectors are shown: $Q_{0}=1$ and $Q_{0}=-1$, with $Q_{\mu}=1$ for $\mu \neq 0$. The data points for the latter are shifted by $0.0025$ along the $x$-axis in order to increase the plot readability.
  \label{fig: plaquettes and cubes}}
\end{figure}

In this section we discuss two local observables: plaquettes and cubes
\begin{align}
    F &=  \Big| \frac{1}{6 V} \sum_{x,y,z,t} \sum_{\mu < \nu} f_{\mu \nu}(x,y,z,t) \Big|, \label{eq:average plaquette}\\
    G &=  \Big| \frac{1}{4 V} \sum_{x,y,z,t} \sum_{\mu < \nu < \rho} g_{\mu \nu \rho}(x,y,z,t) \Big|. \label{eq:average cubes}
\end{align}
According to the factorization theorem \eqref{eq:factorization2}, we expect that $\langle F \rangle$ does not depend on $J_2$ and $\langle G \rangle$ does not depend on $J_1$. Our first numerical results confirm these conclusions. In Figure \ref{fig: plaquettes and cubes} we show the average values $\langle F \rangle$ and $\langle G \rangle$ as functions of $J_1$ and $J_2$ separately. Plots of data obtained in different topological sectors are also indistinguishable, up to statistical uncertainties.

% , but also as a function of the topological sector.

%Note that plaquettes are constructed exclusively from the link variables $m_{\mu}(x,y,z,t)$, whereas the cube observable involves both degrees of freedom, link and faces in a combination given by~\eqref{eq:cube_def}. Based on that, naively, one would expect that $\langle P \rangle$ is a function of $J_1$ only, whereas $\langle G \rangle$ is a function of both coupling constants, $J_1$ and $J_2$. However, due to the factorization theorem Eq.~\eqref{eq:factorization}, $\langle G \rangle$ should not depend on $J_1$. 

In Figure~\ref{fig: plaquettes and cubes} we demonstrate the dependence of $\langle F \rangle$ and $\langle G \rangle$ on $J_1$ (two panels in the upper row) and $J_2$ (two panels in the lower row) coupling constants. 
Motivated by our expectations regarding the phase diagram of the system, i.e. existence of four distinct phases, as described in section~\ref{phase_diag}, linked to the corners of the phase space given by $(J_1, J_2)=(0,0)$, $(0,\infty)$, $(\infty, 0)$ and ($\infty, \infty)$, we select values of $J_1$ and $J_2$ representing each phase: 
\begin{align}
    J_1 &= 0.43 \textrm{ or } 0.46, 
    \label{eq: j1 selected values}\\
    J_2 &= 0.10 \textrm{ or } 1.10.
    \label{eq: j2 selected values}
\end{align}
When varying one of the coupling constants we keep the other in one of the two values.

\begin{figure}
  \centering
  \includegraphics[width=0.345\textwidth, angle=270]{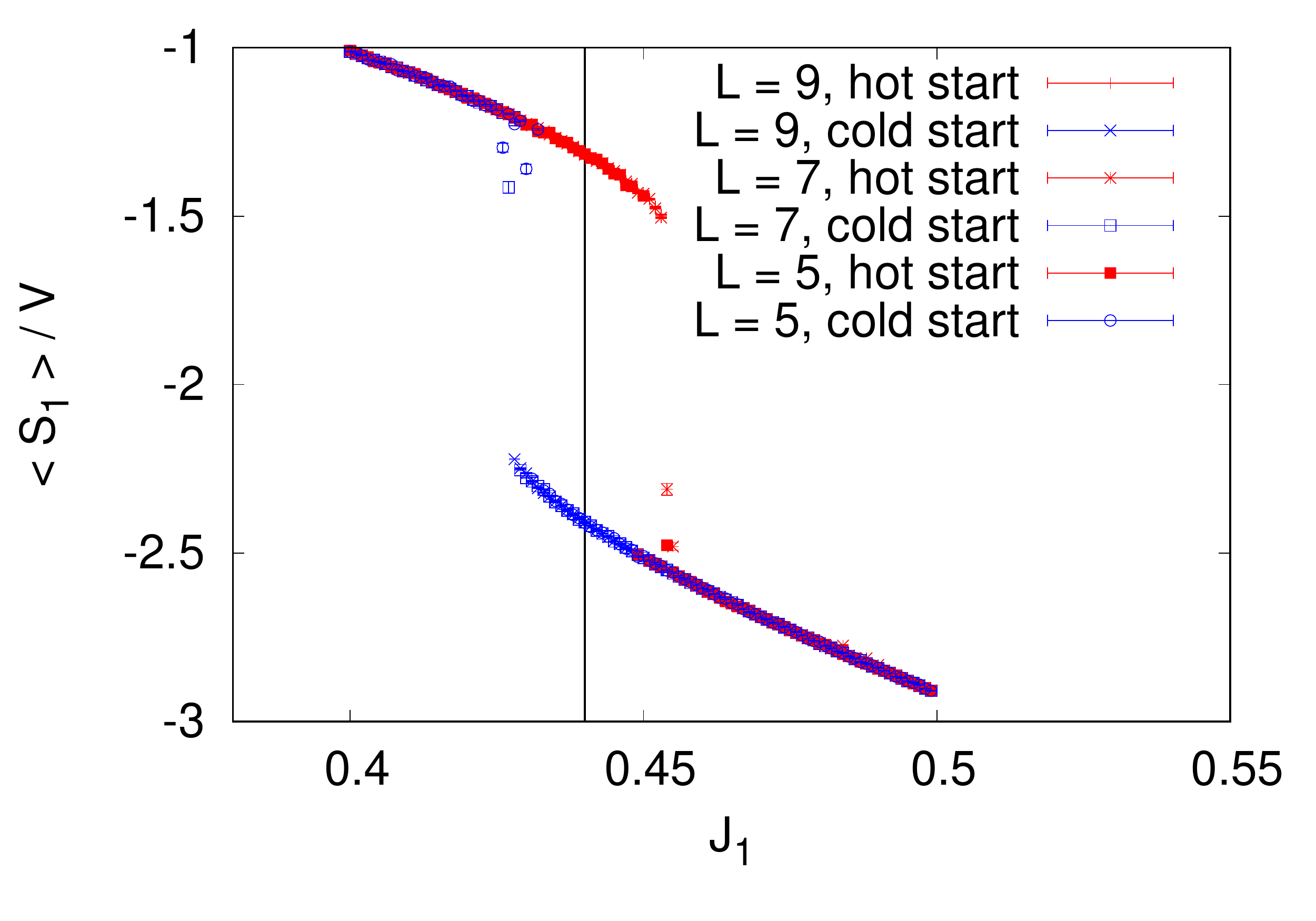}
  \includegraphics[width=0.345\textwidth, angle=270]{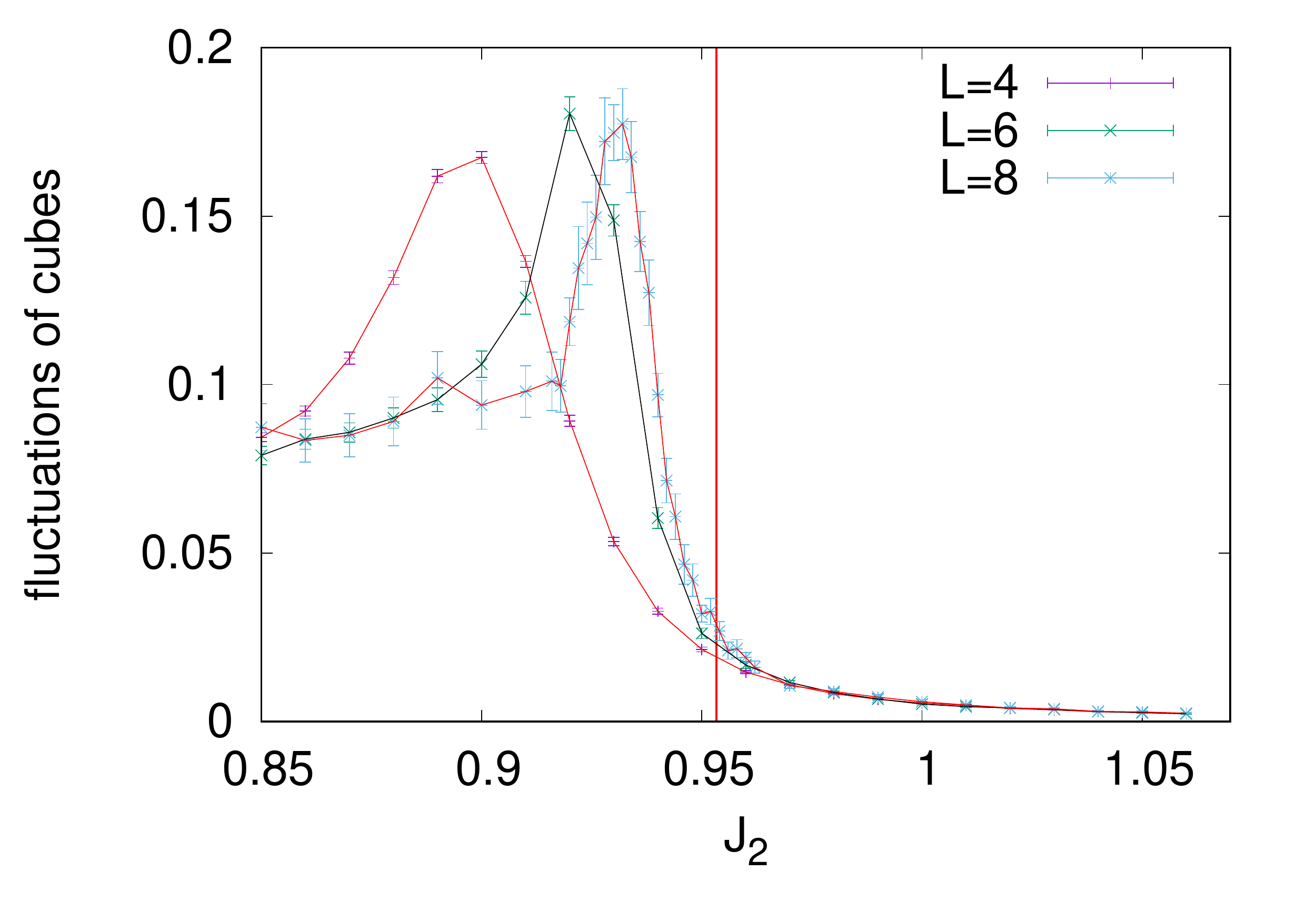}  
  \caption{Left: Results for the action around $J_1$ phase transition. There is a region of $J_1$ couplings where the simulations starting from different initial configurations: cold or hot converge to different local, meta-stable states. Outside of that region, the action has only one minimum and both simulations give the same average value of the plaquette action. Right: Evidence for a second order phase transition in the $J_2$ coupling. Figure shows the fluctuations of the $ G$ observable for simulations at different linear size extends ranging from $L=4$ up to $L=8$. Data points shown are averages of independent simulations conducted in the $Q_0=1$ and $Q_0=-1$ topological sectors. The maximum in the fluctuations approaches the theoretical, infinite limit value shown as the vertical line at $J_2^{\textrm{crit}}$ as discussed around \eqref{eq. j2 critical}. 
  \label{fig:fluctuations and histeresis}}
\end{figure}

We clearly see in Figure \ref{fig: plaquettes and cubes} that $\langle G \rangle$ does not depend on $J_1$, i.e. the values are constant and compatible within their statistical uncertainties for the entire range of $J_1$ values investigated. Similarly, $\langle F \rangle$ does not depend on $J_2$. Near the location of the expected first order phase transition in $J_1$, value of $\langle F \rangle$ drops significantly. We demonstrate the nature of this phase transition in the left panel of Figure \ref{fig:fluctuations and histeresis}. The panel reproduces the results from \cite{creutz_Z2}, where the hysteresis in the average plaquette action in the four-dimensional Wegner model \cite{wegner} was interpreted as a clear sign of a first order phase transition.

As far as $\langle G \rangle$ is concerned, the lower right panel shows a rather smooth dependence. The part of the action proportional to $J_2$ is a function of $\langle G \rangle$, hence we conclude that also the action itself has a continuous dependence on $J_2$. This is in agreement with the expected nature of the phase transition in $J_2$ being second order. We corroborate this with the results shown in the right panel of Figure \ref{fig:fluctuations and histeresis}, where fluctuations of $\langle G \rangle$ are shown to exhibit a drastic change around $J^{\textrm{crit}}_2$. To be precise, we plot $\sqrt{ \langle (G - \langle G \rangle)^2\rangle  V^{-1}}$. Again, all results for $\langle G \rangle$ show no dependence on the change in the $J_1$ coupling constant.

The results shown in Figure~\ref{fig: plaquettes and cubes} provide an illustration of the factorization theorem. Moreover, they support the expected existence of four phases at the four corners of the phase diagram. The more detailed results shown in Figure~\ref{fig:fluctuations and histeresis} suggest that the location of the critical couplings where the phase transitions occur, agree within the accuracy of our simulations with the predictions \eqref{eq:j1 critical value} and \eqref{eq:j2 critical value}. Hence, already the simple, local observables such as $F$ and $G$ provide valuable information about the system. We now turn our attention to non-local observables: Polyakov line and Polyakov planes. The latter, as opposed to the former, are not subject to the factorization theorem and hence are expected to have a non-trivial dependence on both $J_1$ and $J_2$.

% approximately correspond

% \begin{figure}
%   \centering
%   \includegraphics[width=0.95\textwidth]{cubes2.pdf}
%   \caption{Results for fluctuations of  $\langle G \rangle$ as a function of the $J_2$ coupling for linear size extend of $L=4$ and $J_1=0.3$. Data corresponding to simulations at two values of the topological charge are shown and complete agreement within the statistical uncertainties is seen.}
%   \label{fig:cubes}
% \end{figure}

\subsection{Numerical results for non-local observables} \label{sec:MC3}

% Results to be discussed:
% \begin{itemize}
%     \item Polyakov loop - transverse volume dependence in both phases (square root scaling and asymptotically constant behaviour, depending on phase) independent of $J_2$ and topological charge
%     \item Polyakov surface: in each of four phases make plots of transverse volume dependence (single $(J_1, J_2)$ but several topological charges on one plot?)
% \end{itemize}

In this section we discuss results for extended observables. We study in details two such observables: the (volume averaged) Polyakov line, $P_0$, winding around the $x$-direction \eqref{eq:pmu_avg} and the Polyakov plane $P_{01}$, winding around the $x$ and $y$ directions \eqref{eq:pmunu_avg}.

%\begin{equation}
%    P_{01} = \Big| \frac{1}{L L_3} \sum_{z,t} P_{01}(z,t) \Big| 
%\end{equation}

Our expectations for these observables in the four possible phases are based on considerations of the system of infinite size in directions perpendicular to the winding directions. We mimic that limit by taking $L_3 \rightarrow \infty$, which is the direction perpendicular to both $P_0$ and $P_{01}$. We discuss our numerical findings below.

\begin{figure}
  \centering
  \includegraphics[width=0.345\textwidth, angle=270]{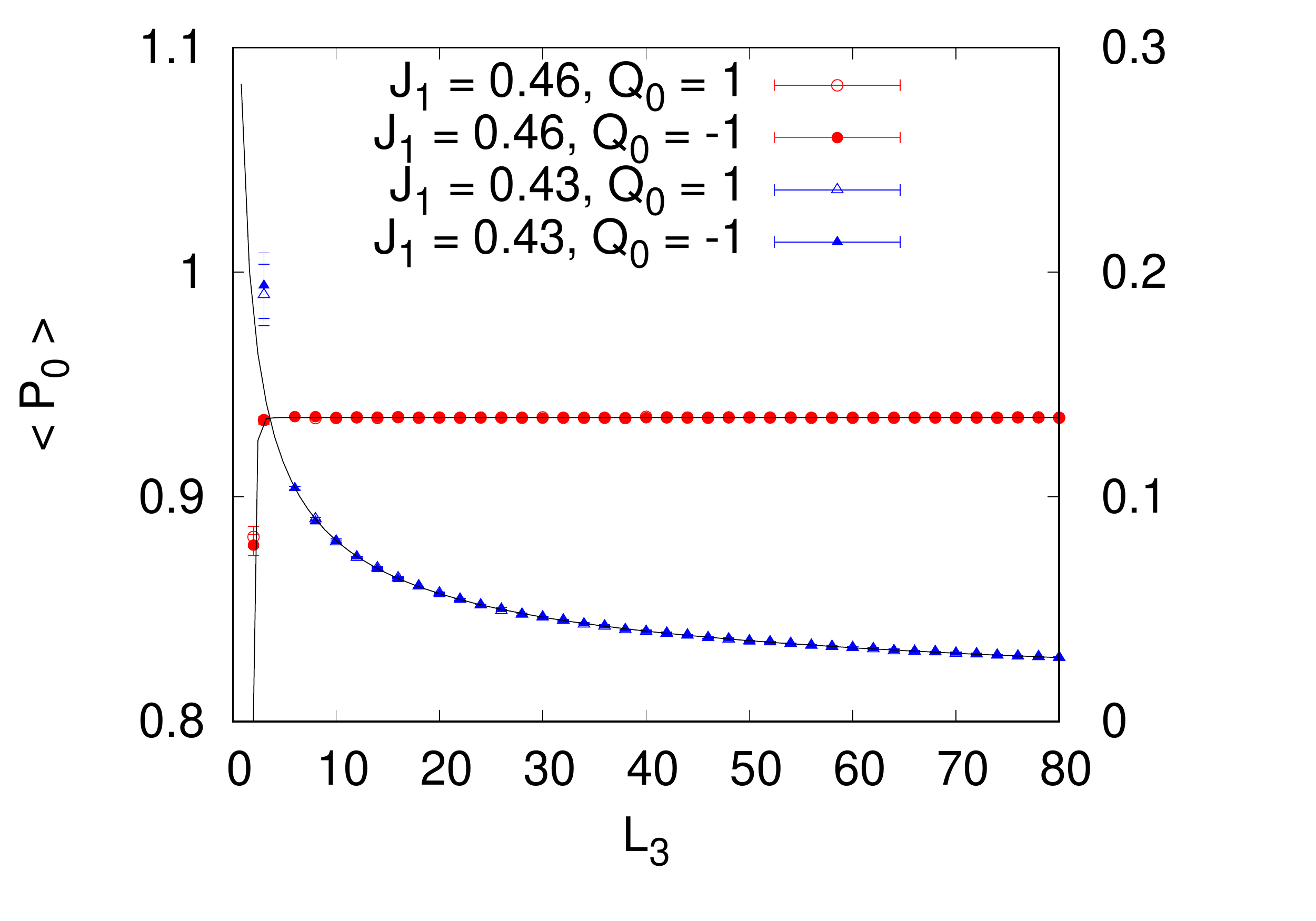}  
  \includegraphics[width=0.345\textwidth, angle=270]{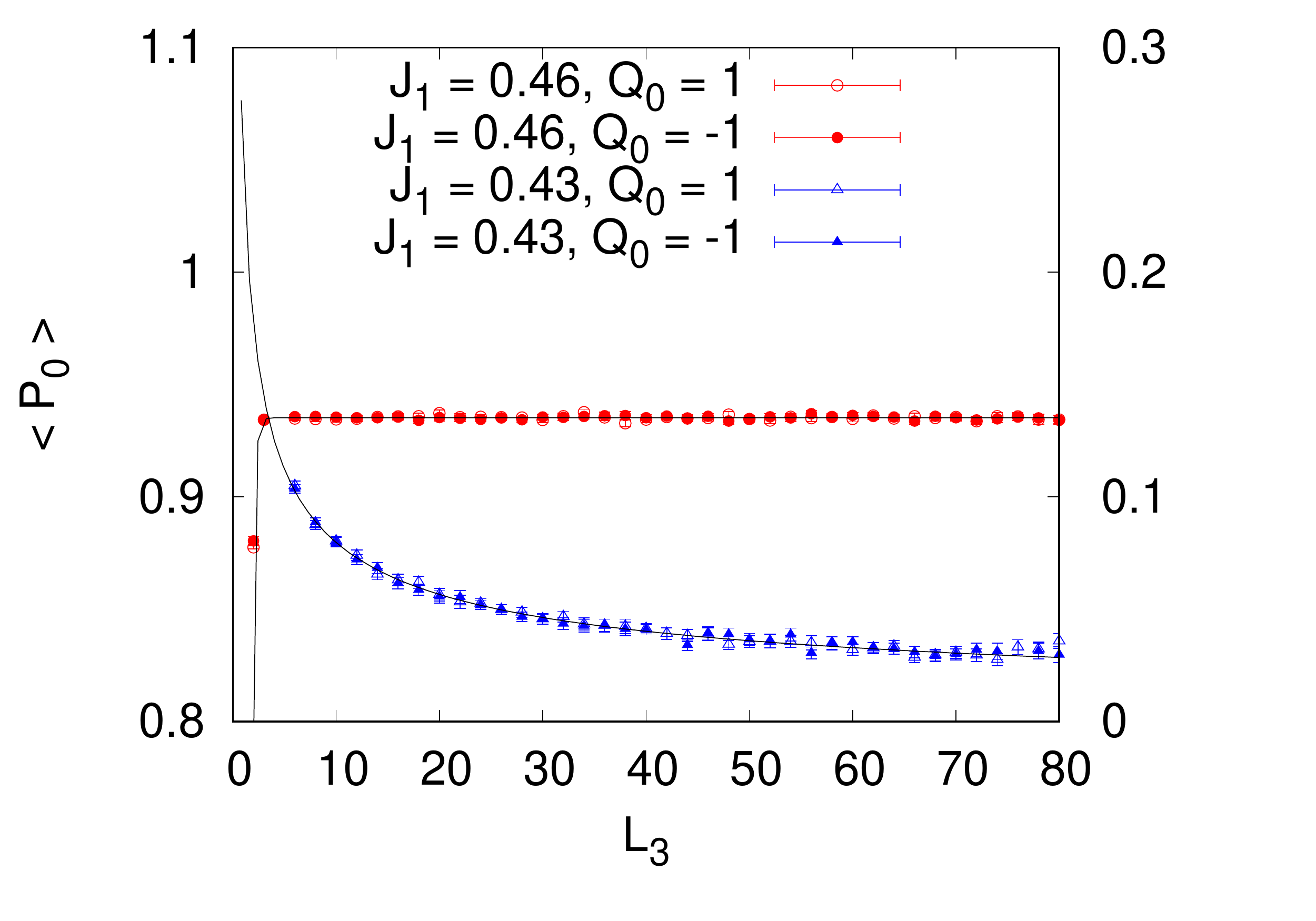}  
  \caption{Demonstration of the dependence of the $\langle P_{0} \rangle$ line on the transverse direction $T$ for small $J_2 = 0.10$ and different $J_1=0.43$ and $J_1=0.46$ for both topological charges. Demonstration of the dependence of the $\langle P_{0} \rangle$ line on the transverse direction $L_3$ for large $J_2 = 1.10$ and different $J_1=0.43$ and $J_1=0.46$ for both topological charges. The left axis shows the values of the data which has a constant nature, whereas the data sets falling towards zero have their values shown on the right axis.}
  \label{fig:plot line}
\end{figure}

\begin{figure}
  \centering
  \includegraphics[width=0.345\textwidth, angle=270]{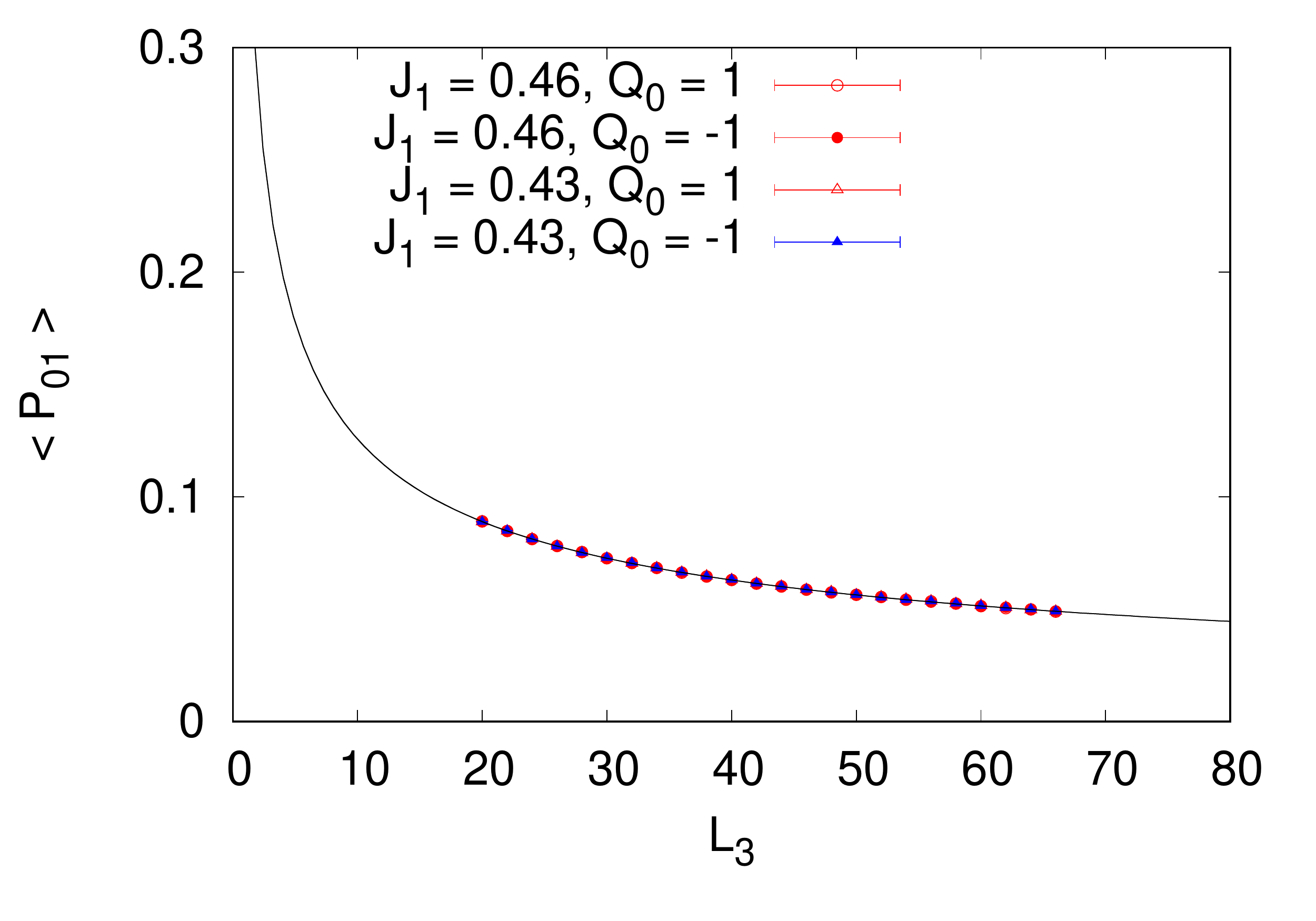}  
  \includegraphics[width=0.345\textwidth, angle=270]{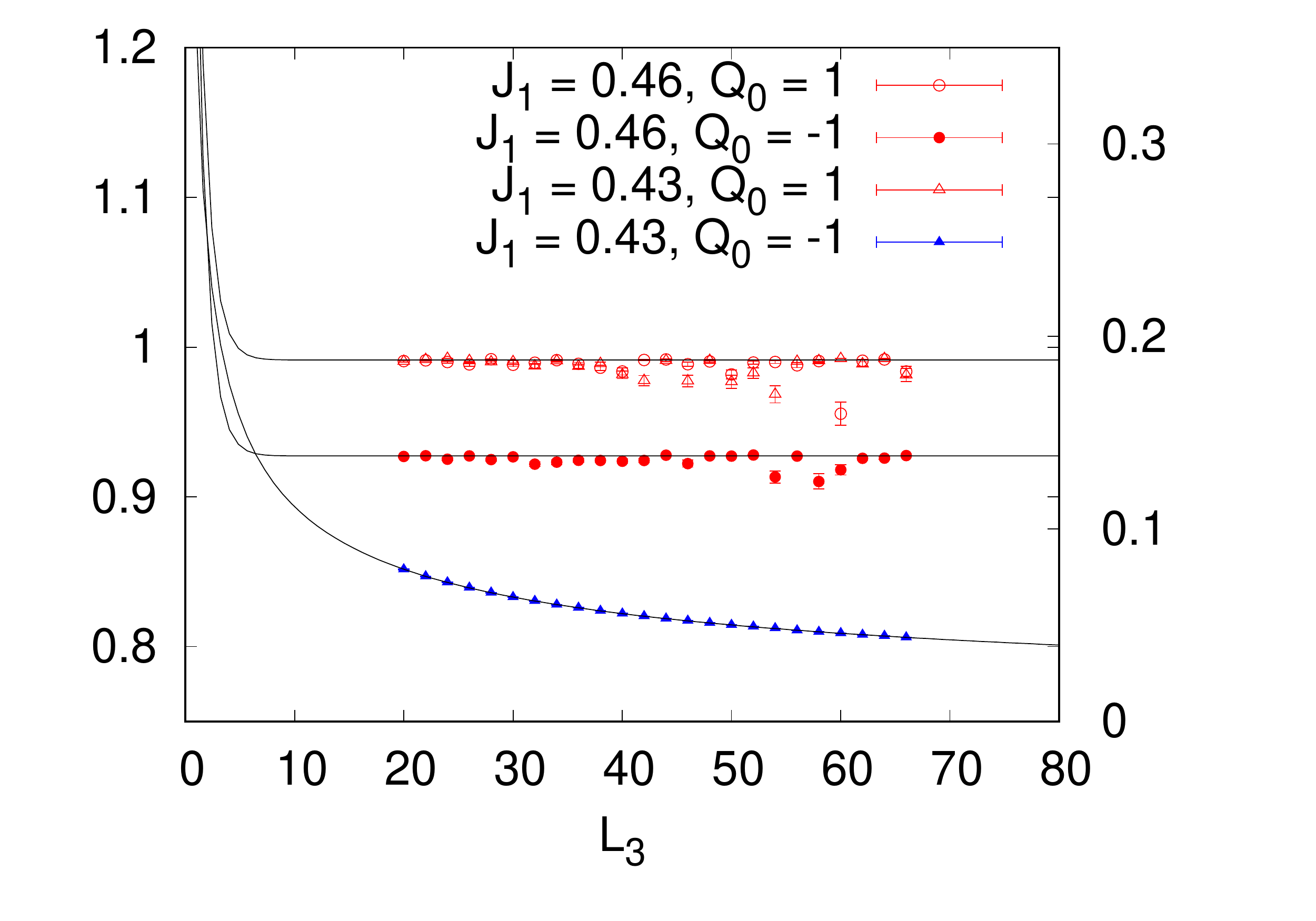}  
  \caption{Demonstration of the $L_3^{-\frac{1}{2}}$ dependence of the $\langle P_{01}\rangle$ plane on the length $L_3$ of one transverse direction for small $J_2 = 0.10$ and different $J_1=0.43$ and $J_1=0.46$ for both topological charges. Demonstration of the dependence of the $\langle P_{01} \rangle$ plane on the transverse direction $L_3$ for large $J_2 = 1.10$ and different $J_1=0.43$ and $J_1=0.46$ for both values of the topological charge. The left axis shows the values of the data which has a constant nature, whereas the data sets falling towards zero have their values shown on the right axis.}
  \label{fig:plot plane}
\end{figure}

We show the numerical results for $\langle P_0 \rangle$ and $\langle P_{01} \rangle$ in Figures \ref{fig:plot line} and \ref{fig:plot plane} at four pairs of coupling constants as a function of the extent of the lattice in the $L_3$ direction. In the left panels we gather results obtained at $J_1 = 0.43$ and $J_1=0.46$ at small $J_2=0.10$, whereas in the right panels we keep the same two values of $J_1$ but we change $J_2$ to a large value, $J_2=1.10$. As~opposed to the previous section, where $\langle F \rangle$ and $ \langle G \rangle$ were discussed as functions of $J_1$ and $J_2$ varying around their critical values, here we study the dependence on the $L_3$ extent at the four values of coupling constants selected in \eqref{eq: j1 selected values} and \eqref{eq: j2 selected values}.

We start with the discussion of Polyakov lines. The observable $\langle P_0 \rangle$ is expected to satisfy the factorization theorem. Indeed, we find that its average value does not depend on the value of the $J_2$ coupling constant. As a consequence, the left and right panels of Figure \ref{fig:plot line}, showing the results for $J_2=0.10$ and $J_2=1.10$ respectively, look very similar. Two scenarios can be realized as the volume of the lattice grows: either the value of $\langle P_0 \rangle $ decreases and ultimately vanishes in the infinite volume limit, or it becomes approximately constant for large volumes. Both scenarios are shown in Figure \ref{fig:plot line}: for $J_1 < J^{\textrm{crit}}_1$ $\langle P_0 \rangle$ decreases as $L_3^{-\frac{1}{2}}$ in the trivial and non-trivial topological sectors. On the contrary, for $J_1 > J^{\textrm{crit}}_1$ we observe that $\langle P_0 \rangle$ stays constant. Data points at very small volumes, $L_3=2$ and $L_3=3$, exhibit finite volume corrections which vanish rapidly with increasing volume. For $L_3 > 4$ a constant fit to the data with $J_1 > J^{\textrm{crit}}_1$ and a fit with an Ansatz of the form $b+c L_3^{-\frac{1}{2}}$ with $b$,$c$ being fit parameters to the data with $J_1 < J^{\textrm{crit}}_1$, describe the data very well within their statistical uncertainties. This allows us to conclude that indeed the Polyakov line is a good order parameters for the phase transition in $J_1$ as it behaves differently on the different sides of $J_1^{\textrm{crit}}$,
\begin{align}
    \langle P_0 \rangle = 0 &\textrm{ for } J_1 < J^{\textrm{crit}}_1, \textrm{ any } J_2, \textrm{ any } Q_0, \ L_3 \to \infty, \\
    \langle P_0 \rangle > 0 &\textrm{ for } J_1 > J^{\textrm{crit}}_1, \textrm{ any } J_2, \textrm{ any } Q_0, \ L_3 \to \infty.
\end{align}

The situation with the Polyakov plane $P_{01}$ is more complicated, as it depends non-trivially on both $J_1$ and $J_2$. Moreover this dependence is different in different topological sectors. We show the data in Figure \ref{fig:plot plane}. Again, the left panel contains results for $J_2 < J^{\textrm{crit}}_2$ while the right panel for $J_2 > J^{\textrm{crit}}_2$. As opposed to the situation with Polyakov lines, now the plots are no longer similar and there is a nontrivial dependence on $J_2$. On the left panel, i.e. for small $J_2$, all data sets show a $L_3^{-\frac{1}{2}}$ dependence signaling that $\langle P_{01} \rangle$ vanishes in this region of phase space in the infinite volume limit. This happens no matter what value of $J_1$ we chose and in both, trivial and non-trivial topological sectors. The right panel contains data for $J_2 > J^{\textrm{crit}}_2$. Only a single data set, the blue one corresponding to $J_1 < J^{\textrm{crit}}_1$ in the topologically charged sector $Q_0 = -1$, vanishes. In all remaining cases the data show a rather constant value as $L_3$ is increased, suggesting a non-zero value in the infinite volume limit. Looking from another perspective, in the trivial topological sector $Q_0=1$, $\langle P_{01} \rangle$ depends only on $J_2$, it vanishes for $J_2 < J^{\textrm{crit}}_2$ and is nonzero for $J_2 > J^{\textrm{crit}}_2$, irrespective of $J_1$. In the non-trivial topological sector, $\langle P_{01} \rangle$ vanishes in three corners of the phase space, except of the region where both $J_1$ and $J_2$ are large, i.e. $J_1 > J^{\textrm{crit}}_1$ and $J_2 > J^{\textrm{crit}}_2$. Hence, $\langle P_{01} \rangle$ at $Q_0=-1$ is sensitive to both $J_1$ and $J_2$ and provides an order parameter for both phase transitions.

Summarizing, for $\langle P_{01} \rangle$ we have in the limit $L_3 \to \infty$:
\begin{align}
    \langle P_{01} \rangle = 0 &\textrm{ for } \textrm{ any } J_1, \ J_2 < J^{\textrm{crit}}_2, \textrm{ any } Q_0,\\
    \langle P_{01} \rangle = 0 &\textrm{ for } J_1 < J^{\textrm{crit}}_1, \ J_2 > J^{\textrm{crit}}_2, \ Q_0 = -1,\\
    \langle P_{01} \rangle > 0 &\textrm{ for } J_1 < J^{\textrm{crit}}_1, J_2 > J^{\textrm{crit}}_2, Q_0 = 1, \\
    \langle P_{01} \rangle > 0 &\textrm{ for } J_1 > J^{\textrm{crit}}_1, J_2 > J^{\textrm{crit}}_2, \textrm{ any } Q_0.
\end{align}

Distinction between phases is seen also by comparing values of $\langle P_{01} \rangle$ for different coupling constants at one finite value of $L_3$, see Table \ref{tab:polyakov plane values}.

\begin{table}[]
    \centering
    \begin{tabular}{|c|c|c||c|}
         \hline
         $J_1$ & $J_2$ & $Q_0$ & $\langle P_{01} \rangle$ \\
         \hline
         0.43  & 0.1 & 1  & 0.0631(2)\\
         0.43  & 0.1 & -1 & 0.0631(2)\\
         \hline
         %0.46  & 0.1 & 1  & 0.062966(15)\\
         %0.46  & 0.1 & -1 & 0.063018(15)\\
         0.46  & 0.1 & 1  & 0.0630(1)\\
         0.46  & 0.1 & -1 & 0.0630(1)\\
         \hline
         0.43  & 1.1 & 1  & 0.9815(3)\\
         0.43  & 1.1 & -1 & 0.0720(2)\\
         \hline
         0.46  & 1.1 & 1  & 0.9838(2)\\
         0.46  & 1.1 & -1 & 0.9238(1)\\
         \hline
    \end{tabular}
    \caption{Assembled average values of $\langle P_{01}\rangle$ in the four regions of phase diagram estimated on a lattice with $L=4$ and $L_3 = 40$.}
    \label{tab:polyakov plane values}
\end{table}

\section{Summary and conclusions}

We have presented an explicit construction of a dynamical lattice model with a local symmetry based on a~2-group. It depends on two coupling constants $J_1, J_2$. We have analyzed the parameter space, first by using dualities to known simpler models, second by simulating the model numerically through Monte Carlo method. Theoretical discussion allows to designate four possible phases in the four corners of the coupling constant plane. In order to study the phase diagram quantitatively, we proposed several candidates for order parameters. Two proposals based on local observables, the average plaquette $F$ and the average cube $G$, are sensitive to the phase transition only in one of the coupling constants. It follows from the factorization theorem, which we formulate and prove, that $F$ constructed from link variables shows the phase transition in $J_1$, whereas $G$ built out of faces shows the phase transition in~$J_2$. The other two candidates for order parameters are non-local observables. Polyakov lines, which are products of link variables, again, feel only the phase transition in the $J_1$ coupling constant. Finally, the Polyakov plane exhibits a non-trivial dependence on both $J_1$ and $J_2$ and hence can be used as an order parameter for both phase transitions. Furthermore, its expectation value depends on the topological charge sector. 

We would like to close this work by mentioning three problems for future study. Firstly, different techniques are required to perform averaging with respect to topological charge sectors. This is because Monte Carlo simulations performed in a fixed topological charge sector do not provide values of weights (partition functions) of distinct sectors. This difficulty is relevant only for those observables for which the average obtained in different topological charge sectors do not agree. The only observable with this property studied in this work is the Polyakov plane. Secondly, it would be interesting to obtain some results about extended surface observables on lattices of topology different than torus, perhaps also for more general crossed modules. Another intriguing question is whether there exists some natural construction of a dynamical higher gauge theory in which factorization theorem does not hold. 

\appendix 

\section{Non-spherical Wilson surfaces} \label{app:surface}

In this appendix we use terminology and notations from \cite{bhr}. Thus in contrast to the remainder of the paper, this part is not fully self-contained.

We consider field configurations on a connected CW-complex $X$ valued in a crossed module $\mathbb G = (\mathcal E, \Phi, \Delta, \rhd)$. They are described by homomorphisms $\Pi_2(X_2, X_1; X_0) \to \mathbb G$, resp. $\Pi_2(X, X_1; X_0) \to \mathbb G$ under the flatness constraint which is the minimization condition for the action $S_2$ from this paper. Replacing $X_2$ by $X$ in the former case and choosing a base point $\ast \in X_0$, we are led to considering homomorphisms $\Pi_2(X, X_1; \ast) \to \mathbb G$. Given such a~homomorphism, we obtain a commutative diagram of group homomorphisms
\[
\begin{tikzcd}
\Phi \arrow[swap]{d}{\Delta} & \pi_2(X,X_1, *) \arrow{r}{h_2} \arrow[swap]{l}{\varphi} \arrow[swap]{d}{\partial} & H_2(X,X_1) \arrow{d}{\partial} \\
\mathcal E &  \pi_1(X_1,*) \arrow{r}{h_1} \arrow[swap]{l}{\epsilon} & H_1(X_1)
\end{tikzcd}
\]
in which $h_i$ are the Hurewicz homomorphisms. Hurewicz theorem and its relative version imply that $h_i$ are surjective with $\ker(h_1)$ and $\ker(h_2)$ generated by expression of the form $\{ \gamma_1 \gamma_2 \gamma_1^{-1} \gamma_2^{-1} \}_{\gamma_1, \gamma_2 \in \pi_1(X_1, \ast)}$ and $\{ (\gamma \rhd \sigma) \sigma^{-1}  \}_{\substack{\gamma \in \pi_1(X_1, \ast) \\ \sigma \in \pi_2(X,X_1,\ast)}}$, respectively. 

Now let $\sigma \in \pi_2(X,X_1, \ast)$ be an element such that $\partial(h_2(\sigma))=0$, i.e.\ such that the relative chain associated to $\sigma$ is a cycle. Since $\partial \circ h_2 = h_1 \circ \partial$, we then have that $\partial \sigma$ belongs to $\ker(h_1)$. It follows that $\varphi_{\sigma}$ belongs to the intersection of $\mathrm{im}(\Delta)$ and the commutant $[\mathcal E, \mathcal E]$ of $\mathcal E$. If this intersection is trivial (e.g. if $\mathcal E$ is abelian, which is satisfied by the crossed module featuring in the model considered in this paper), then $\epsilon_{\partial \sigma}=1$, so $\varphi_{\sigma} \in \ker(\Delta)$. Under a gauge transformation
\begin{equation}
\varphi_{\sigma} \mapsto \xi_{b(\sigma)} \rhd \left( \psi_{\partial \sigma}^{(\epsilon)} \, \varphi_{\sigma}  \right).
\end{equation}
If $\mathcal E$ acts trivially on $\ker(\Delta)$, factor $\xi_{b(\sigma)}$ may be omitted. We claim that furthermore $\psi_{\partial \sigma}^{(\epsilon)}=1$. Indeed, since all $\psi_e$ are in $\ker(\Delta)$ and $\mathcal E$ acts trivially on $\ker(\Delta)$, all epsilons present in the definition of $\psi_{\partial \sigma}^{(\epsilon)}$ may be omitted. On the other hand, since $\partial \sigma$ is a product of commutators, also $\psi_{\partial \sigma}^{(1)}$ is a product of commutators of elements in $\ker(\Delta)$, hence trivial ($\ker(\Delta)$ being abelian). Therefore under the running assumptions $\varphi_{\sigma}$ is gauge-invariant, so it may be used as an observable.

It is interesting to ask whether $\varphi_{\sigma}$ depends on the choice of $\sigma$ representing the cycle $h_2(\sigma)$. If $\sigma'$ is another representative of the same cycle, then $\sigma' = \sigma \sigma_0$ for some $\sigma_0 \in \ker(h_2)$. Thus $\varphi_{\sigma'} = \varphi_{\sigma} \varphi_{\sigma_0}$. We have to describe $\varphi_{\sigma_0}$. By the characterization of $\ker(h_2)$ given earlier we have that $\sigma_0$ is the product $\prod\limits_{i=1}^n (\gamma_i \rhd \tau_i) \tau_i^{-1}$ for some $\gamma_i \in \pi_1(X_1, *)$ and $\tau_i \in \pi_2(X,X_1,*)$. Thus
\begin{equation}
\varphi_{\sigma_0} = \prod_{i=1}^n (\epsilon_{\gamma_i} \rhd \varphi_{\tau_i}) \varphi_{\tau_i}^{-1}.
\end{equation}
This element is trivial if either of the following two conditions is satisfied:
\begin{itemize}
    \item $\varphi_{\tau_i}$ are in $\ker(\Delta)$, i.e.~$\epsilon_{\partial \tau_i}$ are trivial, 
    \item $\epsilon_{\gamma_i}$ are elements of $\mathcal E$ which act trivially on $\Phi$; if $\mathrm{im}(\Delta)$ acts trivially (which is satisfied in the model discussed in this paper), this is automatically satisfied if $\overline \epsilon$ is trivial.
\end{itemize}
In the language used in the main text, these two conditions correspond to $J_1 = \infty$ and trivial topological charge, respectively. Assuming that one of these conditions holds, we find that $\varphi_{\sigma}$ depends on $\sigma$ only through the corresponding homology class in $H_2(X)$ (respectively $H_2(X_2)$ if we do not assume flatness of $\varphi$).

\end{document}